\documentclass[aps, prd, twocolumn, showpacs, showkeys, preprintnumbers, bibnotes, floatfix, longbibliography, notitlepage, nofootinbib, superscriptaddress, superscriptgaddress]{revtex4-2}

\pdfoutput=1
\usepackage[colorlinks=true,breaklinks=true]{hyperref}
\usepackage{amsmath}
\usepackage{amsfonts}
\usepackage{amssymb}
\usepackage{mathrsfs}
\usepackage{graphicx}
\usepackage{color}
\usepackage[utf8]{inputenc}
\usepackage{amsfonts}
\usepackage{graphicx}
\usepackage{booktabs}
\usepackage{multirow}
\usepackage{makecell}
\usepackage{siunitx}
\usepackage{ragged2e}
\usepackage{rotating}
\usepackage{float}
\usepackage[dvipsnames]{xcolor}
\definecolor{darkred}{rgb}{0.5,0,0}
\definecolor{darkblue}{rgb}{0,0,0.5}
\definecolor{firebrick}{rgb}{0.75,0.125,0.125}
\definecolor{darkgreen}{rgb}{0,0.5,0}
\hypersetup{urlcolor=darkblue,
	    citecolor=darkgreen,
	    linkcolor=firebrick}

\usepackage{orcidlink}
\usepackage{lipsum}
\usepackage[normalem]{ulem}
\usepackage{enumitem}
\usepackage[capitalise]{cleveref}
\usepackage{pifont}
\usepackage{subfigure}
\usepackage{physics}
\usepackage{comment}
\usepackage{mwe}

\long\def\exclude#1{}

\newcommand{\ie}{{\it i.e.}}

\newcommand{\eg}{{\it e.g.}}

\newcommand{\fig}{Fig.}

\newcommand{\Refe}{Ref.}
\newcommand{\Refes}{Refs.}

\newcommand{\figu}[1]{\fig~\ref{fig:#1}}

\graphicspath{{figures/}}

\begin{document}

\title{Measuring neutrino mixing above 1~TeV with astrophysical neutrinos}

\author{Mauricio Bustamante}
\email{mbustamante@nbi.ku.dk}
\affiliation{Niels Bohr International Academy, Niels Bohr Institute,\\University of Copenhagen, 2100 Copenhagen, Denmark}

\author{Qinrui Liu}
\email{qinrui\_liu@sfu.ca}
\affiliation{Department of Physics, Engineering Physics and Astronomy, Queen's University, Kingston ON K7L 3N6, Canada}
\affiliation{Arthur B. McDonald Canadian Astroparticle Physics Research Institute,  Kingston ON K7L 3N6, Canada}
\affiliation{Perimeter Institute for Theoretical Physics, Waterloo ON N2L 2Y5, Canada}
\affiliation{Department of Physics, Simon Fraser University, Burnaby, BC V5A 1S6, Canada}

\author{Gabriela Barenboim}
\email{gabriela.barenboim@uv.es}
\affiliation{Departament de F\'isica Te\'orica and IFIC, Universitat de Val\`encia-CSIC, E-46100, Burjassot, Spain}

\date{\today}

\begin{abstract}
We assess the potential for measuring neutrino mixing parameters at energies above 1~TeV, for the first time, using the flavor composition of TeV--PeV astrophysical neutrinos, \ie, the proportion of $\nu_e$, $\nu_\mu$, and $\nu_\tau$. Today, flavor measurements inferred from the 11.4-year IceCube Medium Energy Starting Events sample are insufficient to constrain the mixing parameters due to limited statistics, challenges in flavor identification, and  uncertainty in neutrino production. Yet, upcoming multi-neutrino-telescope observations---even using only existing telescopes---may achieve sensitivity to $\theta_{23}$ and $\theta_{13}$ when combined with  traditional oscillation experiments. We establish the current status and future prospects for testing the three-flavor mixing framework in the previously unexplored TeV--PeV regime and quantify the minimum detectable size of flavor-modifying beyond-Standard-Model effects, providing a roadmap for high-energy neutrino mixing measurements.
\end{abstract}

\maketitle


\textbf{Introduction.---}Neutrino oscillations---the transformation of one neutrino flavor into another---have been extensively studied using solar, atmospheric, reactor, and accelerator neutrinos with energies from about 10~MeV to 1~TeV. Combined observations from these sources~\cite{deSalas:2020pgw, Capozzi:2021fjo, Esteban:2024eli} support a description of neutrino mixing between three active flavors---electron ($\nu_e$), muon ($\nu_\mu$), and tau neutrinos ($\nu_\tau$)---where oscillations are driven by mass differences~\cite{ParticleDataGroup:2024cfk}. Today, the ``mixing parameters'' that govern oscillations are known to within a few to tens of percent~\cite{deSalas:2020pgw, Capozzi:2021fjo, Esteban:2024eli}, but exclusively from sub-TeV measurements.

Above 1~TeV, the three-flavor mixing paradigm remains largely unexplored. At these energies, new neutrino physics that grows with energy could modify the effective mixing parameters, including via new neutrino interactions, mixing with sterile neutrinos, or violations of fundamental symmetries---signaling the discovery of physics beyond the Standard Model (BSM). Atmospheric neutrinos above TeV energies are abundant but unaffected by oscillations while traversing the Earth, eliminating their sensitivity to standard mixing~\cite{Jung:2001dh, Kajita:2014koa}.

\begin{figure}[t!]
 \centering
 \includegraphics[width=\columnwidth]{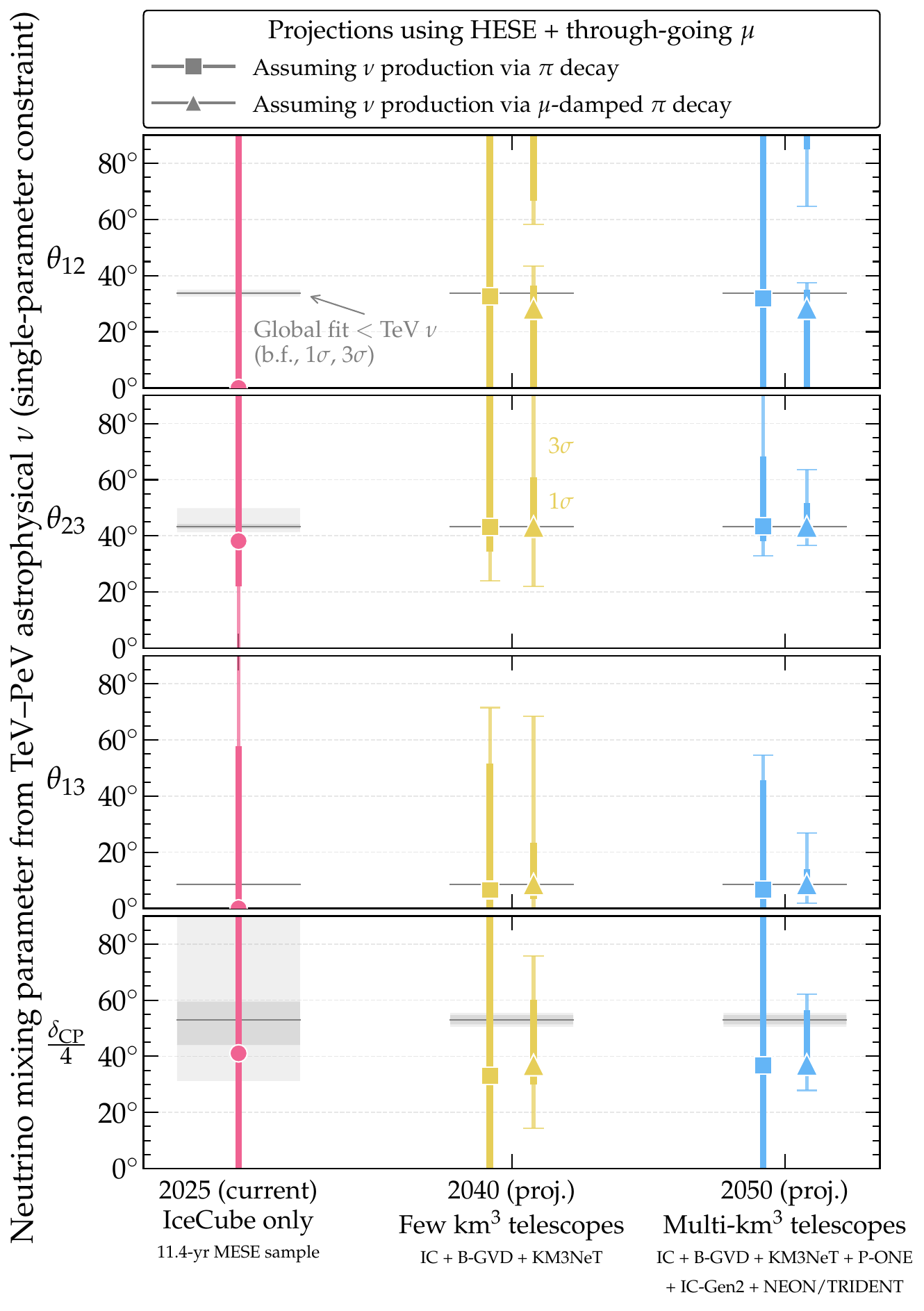}
 \vspace*{-0.6cm}
 \caption{\textbf{\textit{Evolution of measurements of neutrino mixing parameters above 1~TeV.}} Present results are derived from IceCube observations of the flavor composition of high-energy astrophysical neutrinos (11.4-yr MESE sample). Projections (2040/2050) combine HESE and through-going muon samples across multiple detectors under two neutrino production scenarios. Each parameter is constrained individually, profiling over remaining parameters within current global sub-TeV ranges. \textit{Future neutrino telescopes will constrain $\theta_{23}$ and $\theta_{13}$ (and possibly $\delta_{\rm CP}$) above 1~TeV for the first time.}}
 \label{fig:time_evolution}
\end{figure}

High-energy astrophysical neutrinos provide the only probe of both standard and non-standard mixing above 1~TeV. Their potential to do so was recognized early~\cite{Beacom:2003zg, Bhattacharjee:2005nh, Serpico:2005sz, Serpico:2005bs, Balaji:2006wi, Xing:2006xd, Meloni:2006gv, Winter:2006ce, Rodejohann:2006qq, Blum:2007ie, Hwang:2007na, Pakvasa:2007dc, Choubey:2008di, Maltoni:2008jr, Xing:2008fg, Esmaili:2009dz, Meloni:2012nk, Lai:2013isa, Chatterjee:2013tza}, but these analyses preceded or immediately followed their discovery by the IceCube neutrino telescope~\cite{IceCube:2013low}, and were necessarily limited to exploratory studies.

In this work, we quantify how astrophysical neutrinos constrain mixing parameters above 1~TeV. We analyze their TeV--PeV flavor composition---the proportions of $\nu_e$, $\nu_\mu$, and $\nu_\tau$---measured by IceCube and develop projections for future detectors. The flavor composition is a rich probe of astrophysics~\cite{Rachen:1998fd, Athar:2000yw, Crocker:2001zs, Barenboim:2003jm, Beacom:2003nh, Beacom:2004jb, Kashti:2005qa, Mena:2006eq, Kachelriess:2006ksy, Lipari:2007su, Esmaili:2009dz, Choubey:2009jq, Hummer:2010ai, Winter:2013cla, Palladino:2015zua, Bustamante:2015waa, Biehl:2016psj, Bustamante:2019sdb, Ackermann:2019ows, Bustamante:2020bxp, Song:2020nfh, Bhattacharya:2023mmp, Telalovic:2023tcb, Dev:2023znd} and fundamental physics~\cite{Beacom:2002vi, Barenboim:2003jm, Beacom:2003nh, Beacom:2003eu, Beacom:2003zg, Serpico:2005bs, Mena:2006eq, Lipari:2007su, Pakvasa:2007dc, Esmaili:2009dz, Choubey:2009jq, Esmaili:2009fk, Bhattacharya:2009tx, Bhattacharya:2010xj, Bustamante:2010nq, Mehta:2011qb, Baerwald:2012kc, Fu:2012zr, Pakvasa:2012db, Chatterjee:2013tza, Xu:2014via, Aeikens:2014yga, Arguelles:2015dca, Bustamante:2015waa, Pagliaroli:2015rca, Shoemaker:2015qul, deSalas:2016svi, Gonzalez-Garcia:2016gpq, Bustamante:2016ciw, Rasmussen:2017ert, Dey:2017ede, Bustamante:2018mzu, Farzan:2018pnk, Ahlers:2018yom, Brdar:2018tce, Palladino:2019pid, Ackermann:2019cxh, Arguelles:2019rbn, Ahlers:2020miq, Karmakar:2020yzn, Fiorillo:2020gsb, Song:2020nfh, Arguelles:2022tki, MammenAbraham:2022xoc, Telalovic:2023tcb, Telalovic:2025xor}. Unlike terrestrial experiments restricted primarily to transitions between two flavors due to shorter baselines, astrophysical neutrinos undergo full three-flavor mixing over cosmological-scale distances, marking the first rigorous test of three-flavor mixing at these energies.

Figure~\ref{fig:time_evolution} shows the evolution of our results. Present IceCube data cannot constrain the mixing parameters---allowed intervals span the entire physically allowed ranges---but projections to 2040 show that currently operating detectors will achieve meaningful sensitivity to $\theta_{23}$ and $\theta_{13}$ (and possibly $\delta_{\rm CP}$). Although these constraints cannot compete with the precision of sub-TeV global fits~\cite{deSalas:2020pgw, Capozzi:2021fjo, Esteban:2024eli}, they establish an independent probe of neutrino mixing and of the detectability of BSM physics.

\medskip


\textbf{Producing astrophysical neutrinos.---}In cosmic accelerators---\eg, active galaxies, gamma-ray bursts, supernovae---PeV-scale protons interact with ambient matter~\cite{Margolis:1977wt, Stecker:1978ah, Kelner:2006tc} and radiation~\cite{Stecker:1978ah, Mucke:1999yb, Kelner:2008ke, Hummer:2010vx} to produce pions. Pions decay via $\pi^+ \to \mu^+ + \nu_\mu$, followed by $\mu^+ \to \bar{\nu}_\mu + \nu_e + e^+$ (and their charge-conjugates), yielding TeV--PeV neutrinos with flavor ratios $(f_e, f_\mu, f_\tau)_{\rm S} = \left( \frac{1}{3}, \frac{2}{3}, 0 \right)$ at the sources, where $f_{\alpha, {\rm S}} \in [0, 1]$ is the ratio of $\nu_\alpha + \bar{\nu}_\alpha$ ($\alpha = e, \mu, \tau$) to the total flux. Because neutrino telescopes cannot ordinarily distinguish $\nu$ from $\bar{\nu}$, hereafter $\nu_\alpha$ refers to $\nu_\alpha + \bar{\nu}_\alpha$ unless otherwise indicated. Interactions with matter inside the sources are unlikely to modify the flavor ratios~\cite{Mena:2006eq, Razzaque:2009kq, Sahu:2010ap, Varela:2014mma, Xiao:2015gea} (see, however, \Refe~\cite{Dev:2023znd}).

Alternative production mechanisms are possible (see, \eg, \Refe~\cite{Bustamante:2015waa}): muon-damped scenarios with intense magnetic fields yield $(0, 1, 0)_{\rm S}$, while beta decay in neutron-rich sources yields $(1, 0, 0)_{\rm S}$ (though at lower energies). While flavor ratios likely evolve with energy~\cite{Kashti:2005qa, Lipari:2007su}, resolving this evolution is challenging~\cite{Liu:2023flr}. Therefore, we assume energy-independent flavor ratios, representing energy-averaged values. Likewise, we assume isotropic flavor ratios, since there is presently no indication of flavor anisotropy in the neutrino sky~\cite{Telalovic:2023tcb, Telalovic:2025xor}.

Since $\nu_\tau$ production requires rare mesons~\cite{Athar:2005wg, Farzan:2021gbx}, it is expected to be smaller than 10\% of the total.  Thus, we fix $f_{\tau, {\rm S}} = 0$, following standard practice~\cite{Bustamante:2015waa, Bustamante:2019sdb, Song:2020nfh, Liu:2023flr}. The source composition is thus determined by the $\nu_e$ fraction: $(f_{e, {\rm S}}, 1 - f_{e, {\rm S}}, 0)$. When assessing mixing parameter sensitivity, we consider all possible values of $f_{e, {\rm S}}$ to account for the uncertainty in the neutrino production mechanism.  The Supp.~Mat.~shows that relaxing this constraint does not weaken our conclusions.
 
\medskip


\textbf{Neutrino oscillations.---}A neutrino flavor state is a superposition of mass eigenstates, $\nu_i$, \ie, $\nu_\alpha = \sum_{i=1}^3 U_{\alpha i}^\ast \nu_i$, where $U_{\alpha i}$ are elements of the Pontecorvo-Maki-Nakagawa-Sakata (PMNS) mixing matrix. We parametrize the PMNS matrix using three mixing angles ($\theta_{12}$, $\theta_{23}$, $\theta_{13}$) and one CP-violation phase ($\delta_{\rm CP}$) whose sensitivity we assess using astrophysical neutrinos.

For TeV--PeV neutrinos traveling Mpc--Gpc distances from their astrophysical sources to Earth, oscillation lengths are tiny compared to the baselines, leading to rapid oscillations. This, combined with the spread in baselines and limited energy resolution, makes neutrino telescopes sensitive only to the average $\nu_\alpha \to \nu_\beta$ flavor-transition probabilities, $P_{\alpha\beta} = \sum_i \vert U_{\alpha i} \vert^2 \vert U_{\beta i} \vert^2$.

\medskip

\begin{figure}[t!]
 \centering
 \includegraphics[width=\columnwidth]{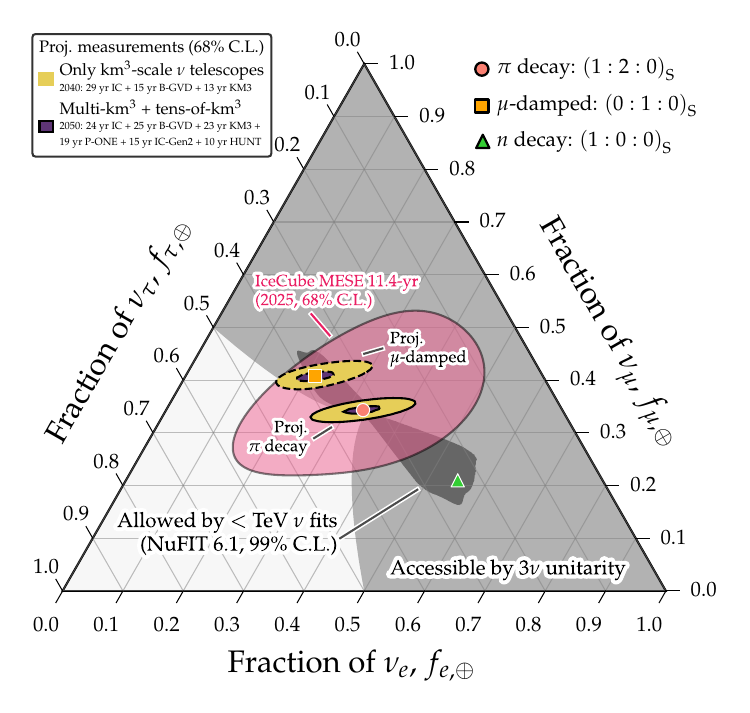}
 \caption{\textbf{\textit{High-energy neutrino flavor composition at Earth.}} Theory-allowed regions vary $f_{e, {\rm S}} \in [0, 1]$ and mixing parameters within experimental uncertainties (``theoretically palatable''~\cite{Bustamante:2015waa, Song:2020nfh}) or unitarity constraints (``unitarity-allowed''~\cite{Ahlers:2018yom}). No $\nu_\tau$ production is assumed. Present IceCube MESE measurements~\cite{Abbasi:2025fjc} have large uncertainty. They are consistent with standard mixing expectations, but cannot constrain mixing parameters. Projected multi-telescope combinations, assuming neutrino production via full pion decay and muon-damped decay, reveal a path to improved precision.}
 \label{fig:flavor_triangle_regions}
\end{figure}

\textbf{Flavor composition at Earth.---}Figure~\ref{fig:flavor_triangle_regions} shows allowed flavor composition at Earth. For a flavor composition at the sources of $(f_e, f_\mu, f_\tau)_{\rm S}$, this is $f_{\alpha, \oplus} = \sum_{\beta} P_{\beta\alpha} f_{\beta, {\rm S}}$. The nominal expectation, for neutrino production via pion decay and assuming known values of the mixing parameters~\cite{Esteban:2024eli}, is approximately $\left( \frac{1}{3}, \frac{1}{3}, \frac{1}{3} \right)_\oplus$. 

The ``theoretically palatable'' region~\cite{Bustamante:2015waa, Song:2020nfh} in \figu{flavor_triangle_regions} varies mixing parameters within current global sub-TeV uncertainties~\cite{Esteban:2024eli} and $f_{e, {\rm S}} \in [0,1]$. The ``unitarity-allowed'' region spans the entire physically allowed ranges of the parameters ($[0^\circ, 90^\circ]$ for $\theta_{12}$, $\theta_{23}$, and $\theta_{13}$, and $[0^\circ, 360^\circ]$ for $\delta_{\rm CP}$) while preserving the unitarity of flavor transitions~\cite{Ahlers:2018yom}. Our predictions of $f_{\alpha, \oplus}$, from which we infer the values of the high-energy mixing parameters, lie within the larger unitarity-allowed region.

The flavor composition offers different sensitivity to different mixing parameters~\cite{Bustamante:2015waa}. As illustration, the long axis of the theoretically palatable region aligns approximately with $\mu$-$\tau$ maximal mixing since $\theta_{23} \approx 45^\circ$~\cite{Esteban:2024eli}. The short-axis width reflects $\theta_{23}$ uncertainty (and, more weakly, $\delta_{\rm CP}$); the long-axis width reflects $\theta_{12}$ uncertainty and the variation of $f_{e, {\rm S}}$. The value of $\theta_{13}$ changes the tilt of the region.  Because flavor measurements are roughly orthogonal to this region (\figu{flavor_triangle_regions}), the sensitivity is primarily to $\theta_{23}$, $\theta_{13}$ (and $f_{e, {\rm S}}$~\cite{Bustamante:2019sdb, Song:2020nfh}).
This differentiated sensitivity carries to the unitarity-allowed region.

\medskip


\textbf{Flavor measurements.---}TeV--PeV neutrino telescopes like IceCube detect Cherenkov light from neutrino interactions, identifying primarily tracks, mostly from $\nu_\mu$, and cascades, mostly from $\nu_e$ and $\nu_\tau$. The latter creates a $\nu_e$-$\nu_\tau$ experimental degeneracy. Additional signatures include double cascades from energetic $\nu_\tau$~\cite{Learned:1994wg} and Glashow resonance enhancement from 6.3-PeV $\bar{\nu}_e$~\cite{Glashow:1960zz, IceCube:2021rpz}.  

Rather than determining the flavors of individual detected events, analyses infer the flavor composition of the diffuse neutrino flux---the sum total of the contributions of all neutrino sources, predominantly unresolved---by comparing the number of detected tracks, cascades, and double cascades~\cite{Mena:2014sja, Palomares-Ruiz:2015mka, Palladino:2015zua, IceCube:2015rro, IceCube:2015gsk, Vincent:2016nut, IceCube:2018pgc, IceCube:2020fpi, Lad:2025fvf, Abbasi:2025fjc}. For our present results, we use the 11.4-year IceCube Medium Energy Starting Events (MESE) flavor composition~\cite{Abbasi:2025fjc} (\figu{flavor_triangle_regions}), the first measurement reporting non-zero content of all flavors at 68\% C.L. MESE events have energies from 1~TeV to 10~PeV and high astrophysical purity~\cite{IceCube:2014rwe, IceCube:2025tgp}. 

For our projections, we follow \Refe~\cite{Liu:2023flr} to infer flavor-composition measurements from a combination of High-Energy Starting Events~\cite{Schonert:2008is, IceCube:2013low, Gaisser:2014bja, IceCube:2014stg, Arguelles:2018awr, IceCube:2020wum} (HESE, tracks, cascades, and double cascades above 60~TeV)---sensitive to all flavors---and through-going muons~\cite{IceCube:2019cia, IceCube:2021xar}, \ie, $\nu_\mu$-initiated tracks born outside the instrumented volume.  We forecast multi-telescope detection in existing IceCube, Baikal-GVD~\cite{Baikal-GVD:2025rhg}, and KM3NeT~\cite{KM3Net:2016zxf}, plus  in future telescopes~\cite{MammenAbraham:2022xoc, Ackermann:2022rqc} P-ONE~\cite{P-ONE:2020ljt}, IceCube-Gen2~\cite{IceCube-Gen2:2020qha}, NEON~\cite{Zhang:2024slv}, TRIDENT~\cite{TRIDENT:2022hql}, and HUNT~\cite{Huang:2023mzt}---up to 30 times the size of IceCube---by scaling IceCube event rates by the detector size, as in \Refe~\cite{Schumacher:2025qca} (also \Refes~\cite{Song:2020nfh, Fiorillo:2022rft, Telalovic:2023tcb, Liu:2023flr, Schumacher:2025qca}).  See the Supp.~Mat.~for detector details.

Figure~\ref{fig:flavor_triangle_regions} illustrates the improvement in flavor-composition measurements.  Projected uncertainties shrink substantially compared to present ones due to increased statistics as multiple detectors accumulate larger event samples. We consider two possible future measurements: one centered on the nominal expectation from full pion decay and one on muon-damped pion decay.

\medskip


\textbf{Parameter inference.---}We adopt a frequentist approach and compute a $\chi^2$ function to assess the sensitivity to the mixing parameters $\boldsymbol{\theta} = (\theta_{12}, \theta_{23}, \theta_{13}, \delta_{\rm CP})$: $\chi_{\rm total}^2 (\boldsymbol{\theta}, f_{e, {\rm S}}) = \chi_{\rm data}^2 (\boldsymbol{\theta}, f_{e, {\rm S}}) + \chi_{\rm prior}^2 (\boldsymbol{\theta}, f_{e, {\rm S}})$, where $\chi_{\rm data}^2(\boldsymbol{\theta}, f_{e, {\rm S}}) = -2 \ln \mathcal{L} (f_{\alpha, \oplus}(\boldsymbol{\theta}, f_{e, {\rm S}}))$, $\mathcal{L}$ is the aforementioned experimental likelihood of flavor-composition measurements, present or future (shown in the Supp.~Mat.), and $\chi_{\rm prior}^2$ represents penalty terms (\textit{pulls}) from sub-TeV oscillation measurements, which affect only the unmeasured parameters.

For present results, the pull terms are $\chi^2$ functions from the NuFIT~6.1~\cite{Esteban:2024eli} global oscillation fit (assuming normal mass ordering with Super-Kamiokande data; other choices change results negligibly). We account for correlations ($\theta_{12}$ vs.~$\theta_{13}$ and $\theta_{23}$ vs.~$\delta_{\rm CP}$), substituting a one-dimensional pull term for the unmeasured partner whenever a parameter is measured.  For projections, the pull terms are Gaussians centered on NuFIT~6.1 best fits with narrower widths from \Refe~\cite{Song:2020nfh}.  In present and projected results, we let $f_{e, {\rm S}}$ float unconstrained in $[0,1]$, reflecting the large uncertainty in neutrino production.

We measure a single parameter of interest at a time, $\mu$, by profiling over all other, nuisance parameters, $\nu$, \ie, $\Delta \chi^2(\mu) = \min_{\nu} [\chi_{\rm total}^2(\mu, \nu)] - \chi^2_\text{global~min}$, and use Wilks' theorem to find confidence intervals.  We find no sensitivity to constrain all parameters simultaneously.
(Bayesian results, seemingly more restrictive, are sensitive to the choice of priors, especially today; see the Supp.~Mat.)

\medskip


\begin{table}[t!]
 \begin{ruledtabular}  
  \caption{\label{tab:results}\textbf{\textit{Sensitivity to neutrino mixing parameters from high-energy astrophysical neutrinos.}} Measurements are derived for each parameter individually while profiling over the others using present (NuFIT 6.1~\cite{Esteban:2024eli}) and future~\cite{Song:2020nfh} sub-TeV global-fit constraints. Present results rely on the 11.4-year IceCube MESE sample~\cite{Abbasi:2025fjc}. Future multi-telescope projections using HESE plus through-going muons consider two production scenarios (full vs.~muon-damped pion decay), showing improved sensitivity to $\theta_{23}$ and $\theta_{13}$ (and $\delta_{\rm CP}$ for the latter). See \figu{results_main} for a comparison with sub-TeV fits and the Supp.~Mat.~for expanded results.} 
  \centering
  \renewcommand{\arraystretch}{1.2}
 \begin{tabular}{lccc}
  \multirow{2}{*}{Parameter} & \multirow{2}{*}{\shortstack{Present \\ (IC MESE 11.4 yr)}} & \multicolumn{2}{c}{Future (HESE + thr.~$\mu$)\footnotemark{}} \\ \cline{3-4}
  & & $\pi$ decay & $\mu$-damped \\
  \hline
  $\theta_{12} [^\circ]$
    & $0.00_{-0.00}^{+90.00}$ & $32.59_{-32.59}^{+57.41}$ & $28.26_{-28.26}^{+6.57}$ \\
  \noalign{\smallskip}
  $\theta_{23} [^\circ]$
    & $38.16_{-16.19}^{+51.84}$ & $43.32_{-4.38}^{+21.61}$ & $43.29_{-1.28}^{+7.33}$ \\
  \noalign{\smallskip}
  $\theta_{13} [^\circ]$
    & $0.00_{-0.00}^{+57.66}$ & $6.67_{-6.67}^{+37.64}$ & $8.63_{-2.06}^{+4.57}$ \\
  \noalign{\smallskip}
  $\delta_{\rm CP} [^\circ]$
    & $164_{-164}^{+196}$ & $147_{-147}^{+213}$ & $148_{-12}^{+76}$ \\
  \noalign{\smallskip}
  $f_{e, {\rm S}}$
    & $0.04_{-0.04}^{+0.49}$ & $0.33_{-0.02}^{+0.02}$ & $0.00_{-0.00}^{+0.03}$ \\
 \end{tabular}
 \end{ruledtabular}  
 \footnotetext[1]{Estimated year 2050, using 24~yr of IceCube (IC) + 25~yr of Baikal-GVD + 23~yr of KM3NeT + 19~yr of P-ONE + 15~yr of IceCube-Gen2 + 10~yr of HUNT.}
\end{table}

\begin{figure}[t!]
 \centering
 \includegraphics[width=\columnwidth]{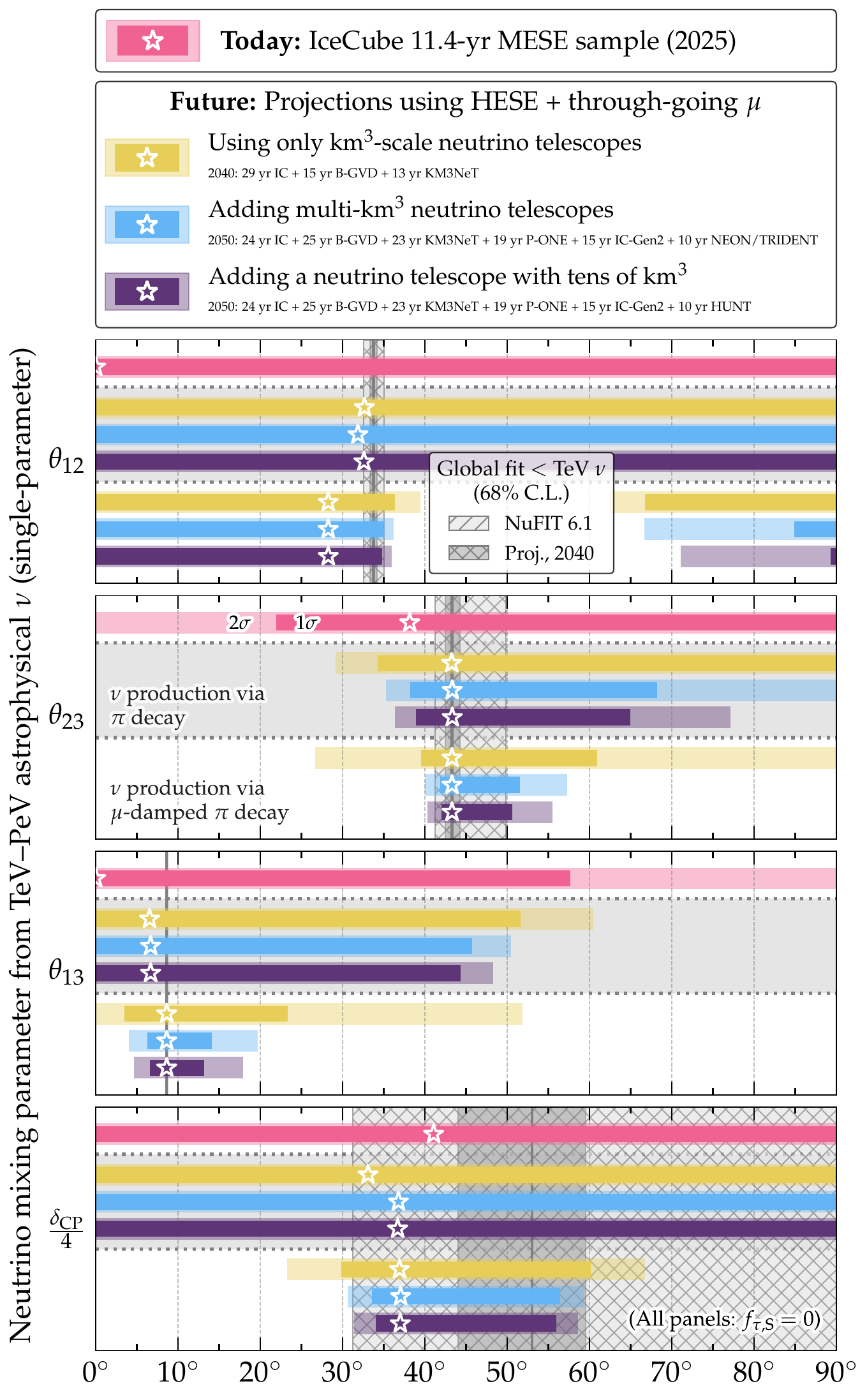}
 \vspace*{-0.5cm}
 \caption{\textbf{\textit{Constraints on neutrino mixing parameters with high-energy astrophysical neutrinos.}} Present constraints from the 11.4-yr IceCube MESE sample~\cite{Abbasi:2025fjc} are consistent with three-flavor mixing but cannot  constrain individual parameters. Projected measurements from multi-telescope observations show  sensitivity to $\theta_{23}$ and $\theta_{13}$ (and $\delta_{\rm CP}$ under muon-damped production). For comparison, we show current~\cite{Esteban:2024eli} and projected~\cite{Song:2020nfh} sub-TeV global-fit ranges. \textit{These results represent the first rigorous assessment of mixing parameter sensitivity at TeV--PeV energies.}}
 \vspace*{-0.5cm}
 \label{fig:results_main}
\end{figure}

\textbf{Current status: IceCube.---}Figure~\ref{fig:results_main} and Table~\ref{tab:results} show the resulting currently allowed ranges of the TeV--PeV mixing parameters, inferred from the 11.4-yr IceCube MESE flavor measurements.  These ranges span nearly the entire physically allowed ranges, indicating insufficient power to meaningfully constrain them. 

This lack of sensitivity arises from three main factors: (1) statistics that, despite reaching about 10,000 MESE events with high astrophysical purity, yield large errors in flavor measurements, (2) challenges in flavor identification, and (3) theoretical uncertainty in the source flavor composition $f_{e, {\rm S}}$, which allows for degeneracies with different mixing parameters.

Even without tight constraints, the consistency of the observed flavor composition with standard expectations delivers physical insight. The data show no evidence for large deviations from three-flavor mixing, ruling out BSM scenarios that yield dramatically different flavor ratios~\cite{Bustamante:2015waa, Rasmussen:2017ert}. This establishes the first consistency check of the mixing framework at TeV energies, even though the mixing parameters cannot yet be constrained.

\medskip


\textbf{Multi-telescope projections.---}Projections using multi-telescope observations show pathways to achieving meaningful sensitivity to selected mixing parameters.

Figure~\ref{fig:results_main} and Table~\ref{tab:results} show projections using HESE plus through-going muon observations from existing (IceCube, Baikal-GVD, KM3NeT) and next-generation detectors (P-ONE, IceCube-Gen2, NEON/TRIDENT, HUNT), accumulated until 2040 and 2050.  Detectors have different sizes and start operations at different times~\cite{Schumacher:2025qca}.  The Supp.~Mat.~contains simulation details.

Our projections reveal future sensitivity to $\theta_{23}$ and $\theta_{13}$, aligned with pioneering work~\cite{Beacom:2003zg, Bhattacharjee:2005nh, Serpico:2005sz, Serpico:2005bs, Balaji:2006wi, Xing:2006xd, Meloni:2006gv, Winter:2006ce, Rodejohann:2006qq, Blum:2007ie, Hwang:2007na, Pakvasa:2007dc, Choubey:2008di, Maltoni:2008jr, Xing:2008fg, Esmaili:2009dz, Meloni:2012nk, Lai:2013isa, Chatterjee:2013tza} but strengthened by improved sub-TeV global fits and realistic simulations.
Assuming production via pion decay, future $2\sigma$ intervals remain one-sided except for $\theta_{23}$ in our farthest projections, reaching 50\% precision on $\theta_{23}$ and factor-of-6 sensitivity on $\theta_{13}$.  Assuming muon-damped production, constraints are a factor-of-3 tighter, reaching 17\% precision on $\theta_{23}$, 53\% on $\theta_{13}$, and 51\% on $\delta_{\rm CP}$.  The weaker results assuming pion decay stem from it yielding about $\left( \frac{1}{3}, \frac{1}{3}, \frac{1}{3} \right)_\oplus$ at Earth which, as explained in Ref.~\cite{Bustamante:2019sdb}, maps back into $\left( \frac{1}{3}, \frac{1}{3}, \frac{1}{3} \right)_{\rm S}$ at the sources regardless of the mixing parameters, diluting the sensitivity to specific values. 

Broadly stated, improving the measurement of the $\nu_\mu$ and $\nu_\tau$ content tightens constraints on $\theta_{23}$ and $\theta_{13}$ (and $\delta_{\rm CP}$ in the muon-damped case). Conversely, improving the $\nu_e$ measurement would nominally tighten $\theta_{12}$. However, in practice, sensitivity to $\theta_{12}$ is washed out by the large uncertainty in the source, $f_{e, {\rm S}}$, which mimics the effect of $\theta_{12}$ on the flavor composition at Earth.

Progress may be accelerated through improvements in flavor identification in water-based detectors (\eg, Baikal-GVD, KM3NeT)---which are less affected than ice-based detectors by light scattering~\cite{KM3Net:2016zxf}---the use of dedicated event templates~\cite{IceCube:2020fpi, Lad:2025fvf}, and novel techniques such as muon and neutron echoes~\cite{Li:2016kra, Steuer:2017tca, Farrag:2023jut, Dutta:2025qgk}, not considered in our projections.  Better understanding of astrophysical sources would tighten constraints on $f_{e, {\rm S}}$. Future flavor measurements using MESE, more abundant than HESE, will further accelerate progress.

\medskip


\textbf{Implications for BSM physics.---}Even in the absence of tight constraints on standard mixing parameters, our analysis establishes important benchmarks for detecting BSM neutrino physics at TeV energies.

The allowed intervals in \figu{results_main} and Table~\ref{tab:results} quantify the minimum size that BSM modifications to the effective mixing parameters must have to be detectable.  At present, given the lack of sensitivity, no size of modifications relative to standard values is detectable.  Future multi-telescope observations will be sensitive to modifications of 15--20$^\circ$ for $\theta_{23}$ and $\theta_{13}$ (and slightly larger ones for $\delta_{\rm CP}$ under muon-damped neutrino production).

By the close of the coming decade, this level of sensitivity will be enough to constrain certain BSM scenarios that posit large modifications to neutrino mixing and flavor ratios at Earth.  This includes, \eg, Lorentz-invariance violation~\cite{Barenboim:2003jm, Arguelles:2015dca, Bustamante:2015waa}, active-sterile neutrino mixing~\cite{Arguelles:2019tum, Ahlers:2020miq}, and new neutrino interactions~\cite{Bustamante:2018mzu}. 

The unique value of TeV--PeV mixing measurements lies in their sensitivity to BSM effects that grow with energy, complementing constraints from lower energies, where these effects would be undetectable. Beyond 100~PeV, BSM effects may become more prominent, enabling their detection via flavor measurements in future ultra-high-energy neutrino telescopes~\cite{Testagrossa:2023ukh, Coleman:2024scd, Cummings:2025tqc}.  

\medskip


\textbf{Conclusions and outlook.---}TeV--PeV astrophysical neutrinos offer the first test of three-flavor mixing above 1~TeV.  Using their flavor composition, we have quantified realistic prospects for measuring neutrino mixing parameters using present data and projections.

Present IceCube data (the 11.4-year MESE sample) leave the parameters unconstrained. However, our projections show improvement when combined with future sub-TeV measurements.  Multi-telescope observations by 2050 will achieve 50\% precision on $\theta_{23}$ and factor-of-6 sensitivity on $\theta_{13}$ if neutrinos are produced via standard pion decay.  If production occurs instead via muon-damped pion decay, precision is boosted to 17\% on $\theta_{23}$, 53\% on $\theta_{13}$, and 51\% on $\delta_{\rm CP}$.
While coarser than sub-TeV global fits, these results establish the first TeV--PeV constraints of the three-flavor paradigm.

Beyond confirming standard mixing at high energies, these measurements establish benchmarks for BSM sensitivity that account for astrophysical unknowns. BSM deviations from sub-TeV parameter values will be detectable at $2\sigma$ if they exceed $10^\circ$--$20^\circ$ in $\theta_{23}$ and $\theta_{13}$.

Progress requires three advances: increased statistics, improved flavor reconstruction, and, crucially, better astrophysical source characterization. The latter could unlock sensitivity to $\theta_{12}$. These developments will transform high-energy astrophysical neutrinos from discovery-era into robust probes of mixing beyond terrestrial reach.

\medskip


\textbf{Acknowledgements.---}The authors are grateful for the contribution of Bernanda Telalovic in the early stages of this work and for the feedback of Subir Sarkar on the manuscript.   MB is supported by {\sc Villum Fonden} under project no.~29388. This work used the Tycho supercomputer hosted at the SCIENCE High Performance Computing Center at the University of Copenhagen.  QL is supported by CFREF and NSERC of Canada through the Arthur B. McDonald Canadian Astroparticle Physics Research Institute. This work used equipment funded by the Canada Foundation for Innovation and the Province of Ontario, and housed at the Queen's Centre for Advanced Computing. Research at Perimeter Institute is supported by the Government of Canada through the Department of Innovation, Science, and Economic Development, and by the Province of Ontario. GB is supported by the Spanish grants CIPROM/2021/054 (Generalitat Valenciana), PID2023-151418NB-I00 funded by MCIU/AEI/10.13039/501100011033/, and by the European ITN project HIDDeN (H2020-MSCA-ITN-2019/860881-HIDDeN).


%


\newpage
\clearpage
\appendix


\onecolumngrid

\begin{center}
 \large
 Supplemental Material for\\
 \smallskip
 {\it Measuring neutrino mixing above 1~TeV with astrophysical neutrinos}
\end{center}

\twocolumngrid


\section{Experimental likelihood of flavor-composition measurements}
\label{app:flavor_likelihood}

\renewcommand{\theequation}{A\arabic{equation}}
\renewcommand{\thefigure}{A\arabic{figure}}
\renewcommand{\thetable}{A\arabic{table}}
\setcounter{figure}{0} 
\setcounter{table}{0} 

This appendix details the construction and properties of the experimental flavor-composition likelihood used throughout our analysis, including present IceCube measurements and future multi-telescope projections. 

In the main text, we used the experimental flavor-composition likelihood, $\mathcal{L}$, to compute the Bayesian posterior of the neutrino mixing parameters.  This likelihood represents how compatible a test flavor composition at Earth, $( f_e, f_\mu, f_\tau )_\oplus$, is with the flavor composition inferred from neutrino-initiated events detected by IceCube or by combinations of neutrinos telescopes.  It also represents the precision with which neutrino telescopes can measure the flavor composition.

All of our projections for the flavor-composition likelihood are computed using the collective detection of HESE plus through-going muons by multiple neutrino telescopes, existing and future, following the methods from \Refe~\cite{Liu:2023flr}.   We do not use MESE in our projections because its detector effective area or detailed detector response to do so are unavailable publicly.

In the projections, we make the simplifying assumption that each neutrino telescope is a scaled version of IceCube, \ie, we scale the event rates of the telescope by its size relative to IceCube, using Table~I in \Refe~\cite{Schumacher:2025qca}.  We estimate the year in which these projections might become possible by adopting the tentative detector start dates from the same table, ignoring contributions of intermediate construction stages.  

The projected flavor measurements are centered either at $(0.32, 0.34, 0.36)_\oplus$, as expected from neutrino production by full pion decay, or at $(0.22, 0.41, 0.37)_\oplus$, as expected from muon-damped pion decay. We assume the astrophysical neutrino energy spectrum is a simple power law $\propto E^{-2.5}$, where $E$ is the neutrino energy, consistent with the IceCube measurement of it combining multiple event samples~\cite{IceCube:2015gsk}. 

\begin{figure}[t!]
 \centering
 \includegraphics[width=0.9\columnwidth]{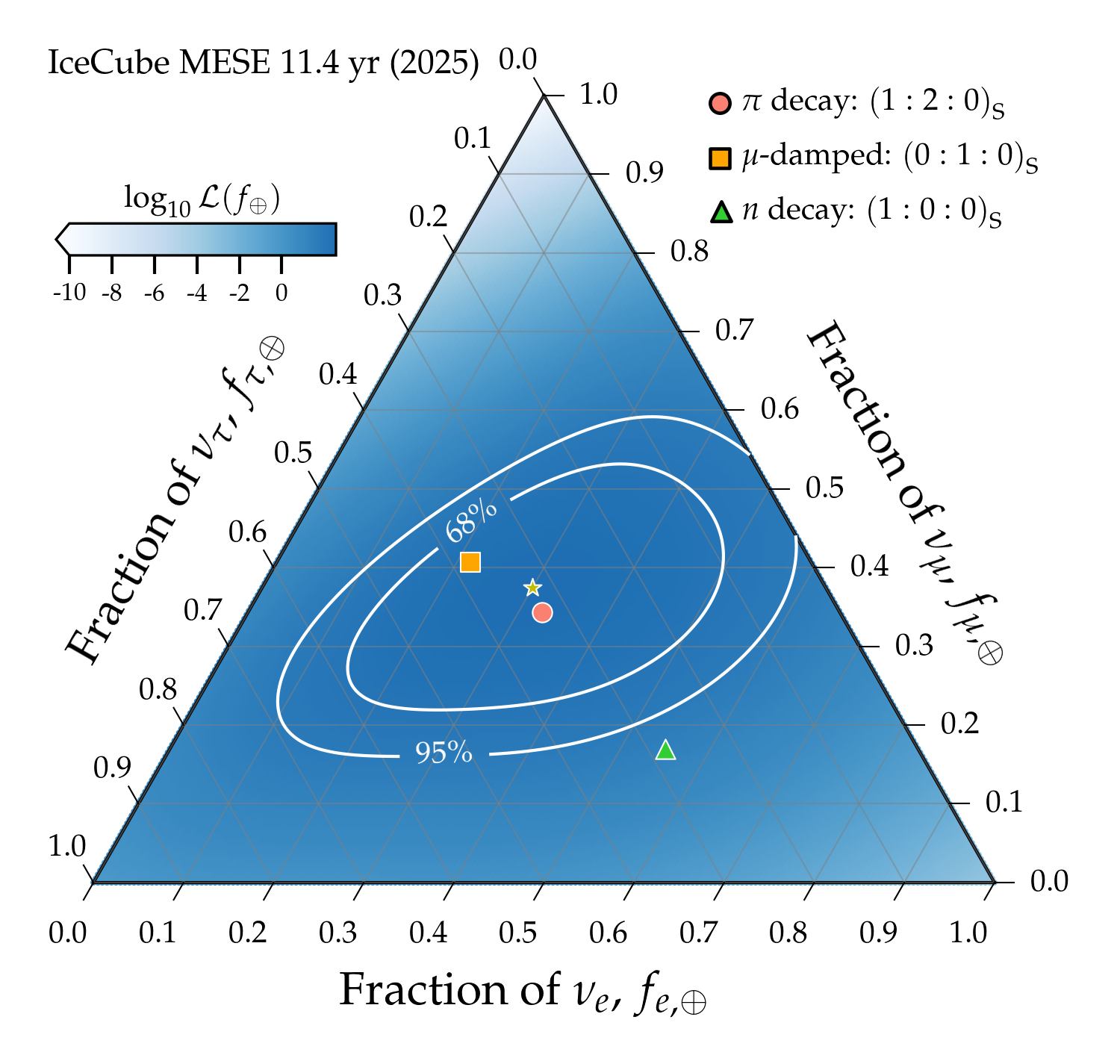}
 \caption{\textbf{\textit{Present experimental likelihood of flavor-composition measurements}}. This likelihood, $\mathcal{L}$, is a parametric form of the allowed contours reported \Refe~\cite{Abbasi:2025fjc}, inferred from the 11.4-yr IceCube MESE sample~\cite{IceCube:2025tgp}.}
 \label{fig:flavor_likelihood_mese}
\end{figure}

\begin{figure*}[t] 
    \centering
    \begin{minipage}[t]{0.48\textwidth}
        \centering
        \includegraphics[width=\linewidth]{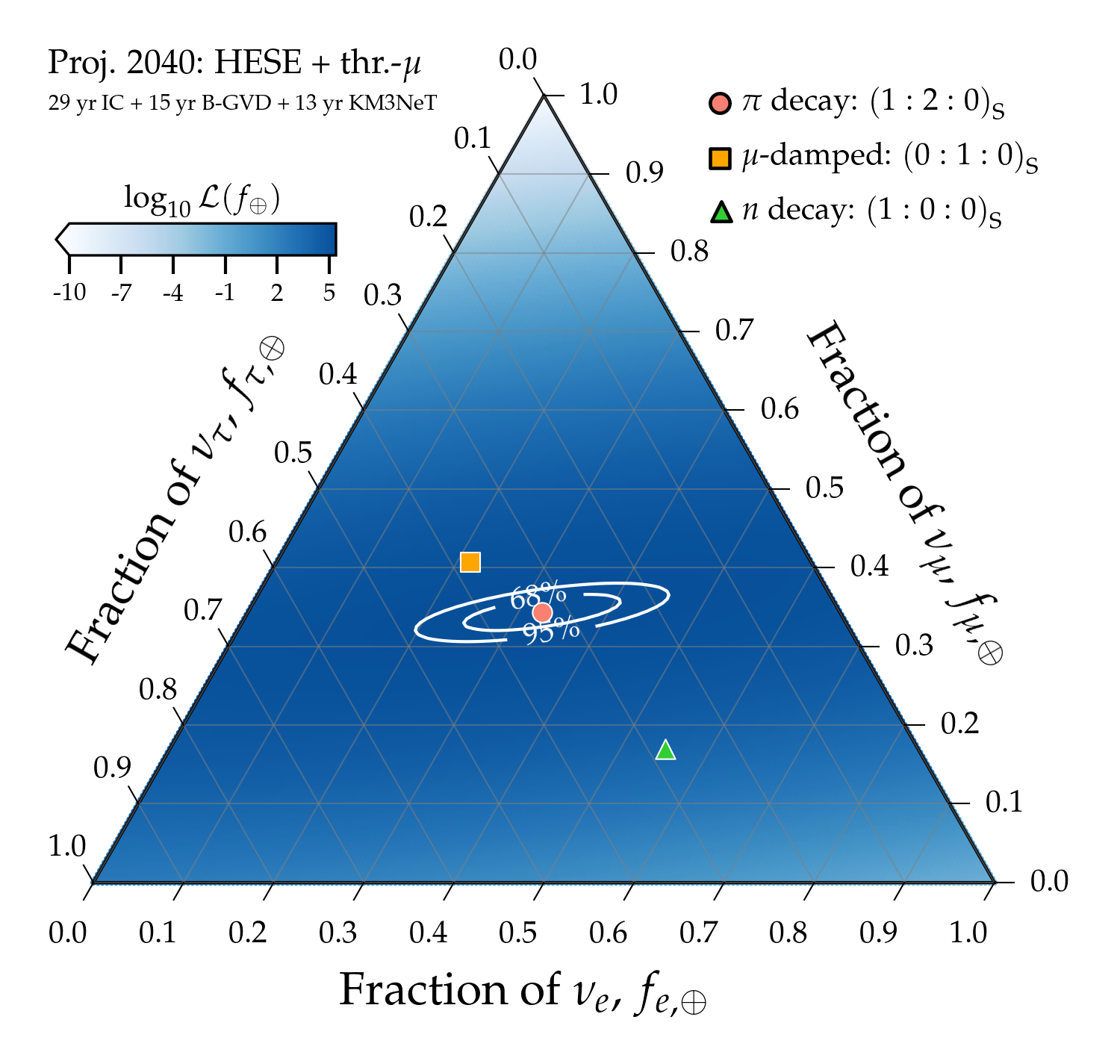}
        \par\vspace{-10pt}
        \includegraphics[width=\linewidth]{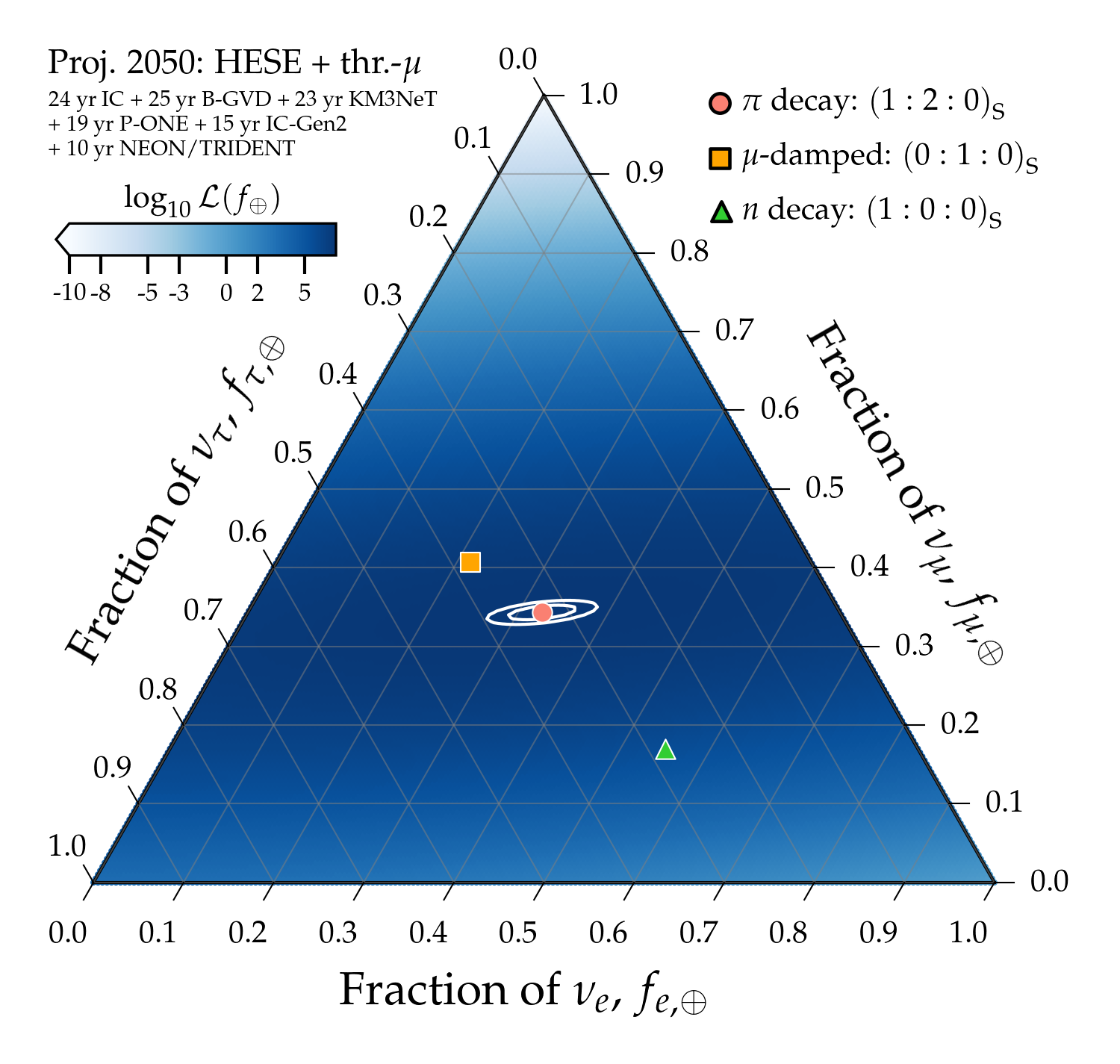}
        \par\vspace{-10pt}
        \includegraphics[width=\linewidth]{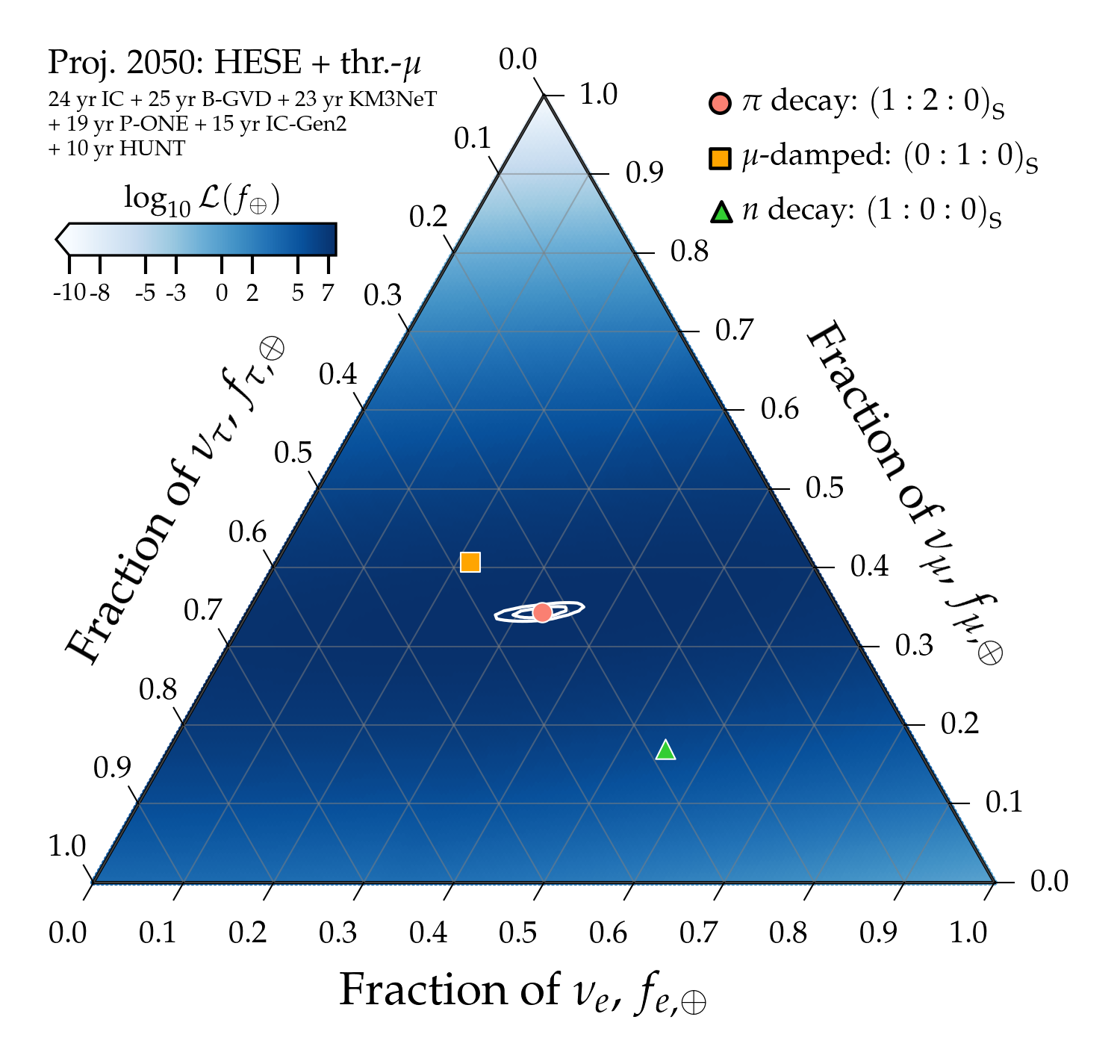}
    \end{minipage}
    \hfill 
    \begin{minipage}[t]{0.48\textwidth}
        \centering
        \includegraphics[width=\linewidth]{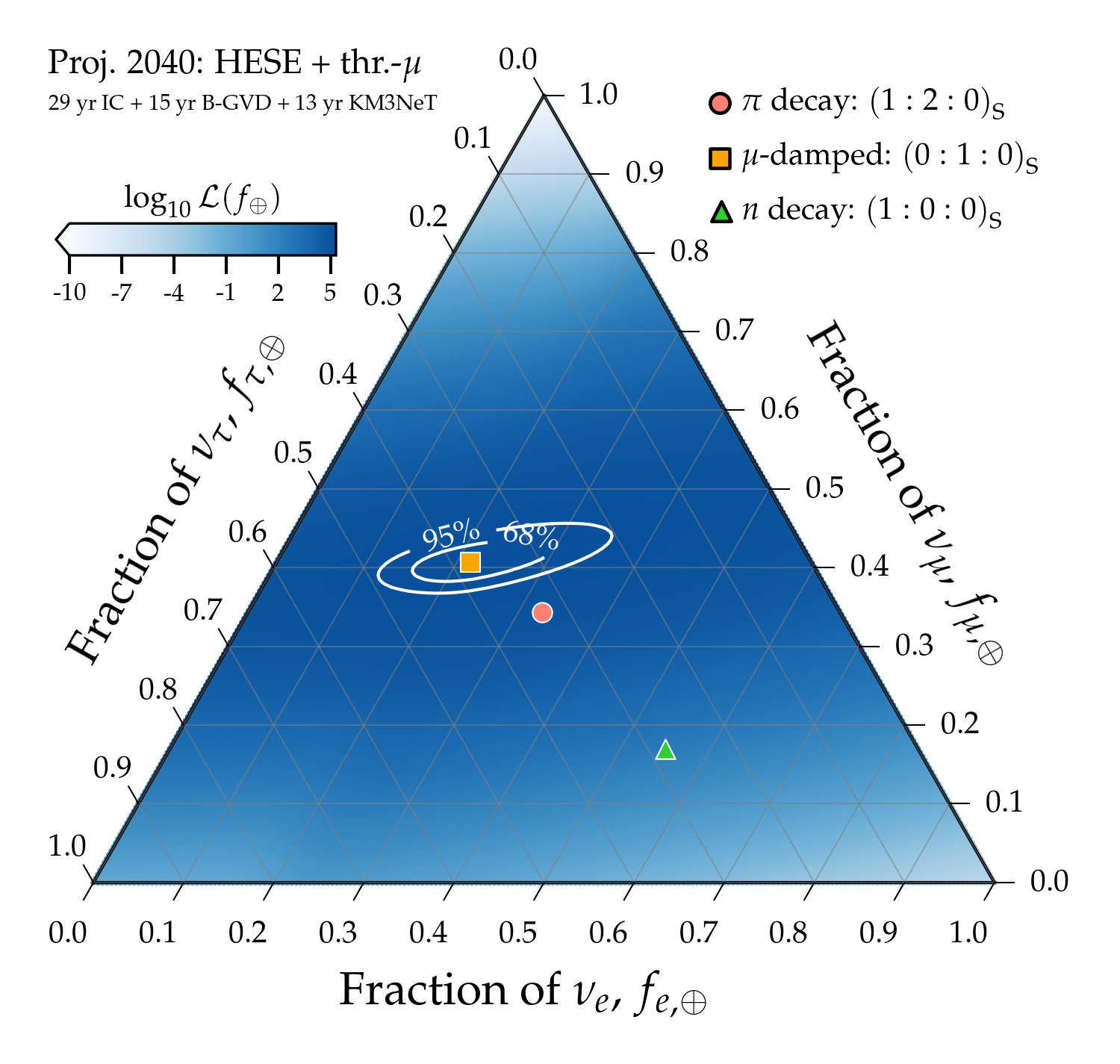}
        \par\vspace{-10pt}
        \includegraphics[width=\linewidth]{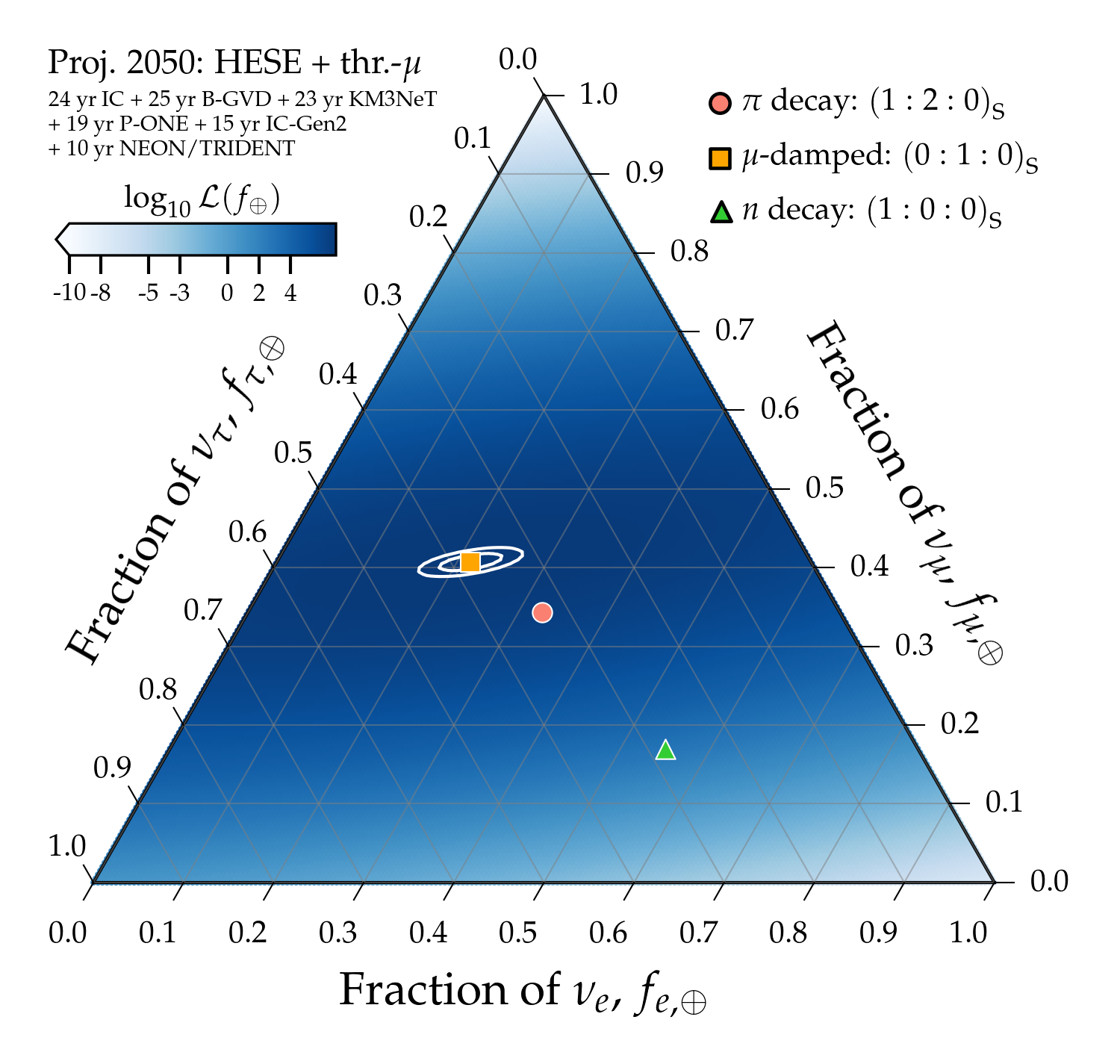}
        \par\vspace{-10pt}
        \includegraphics[width=\linewidth]{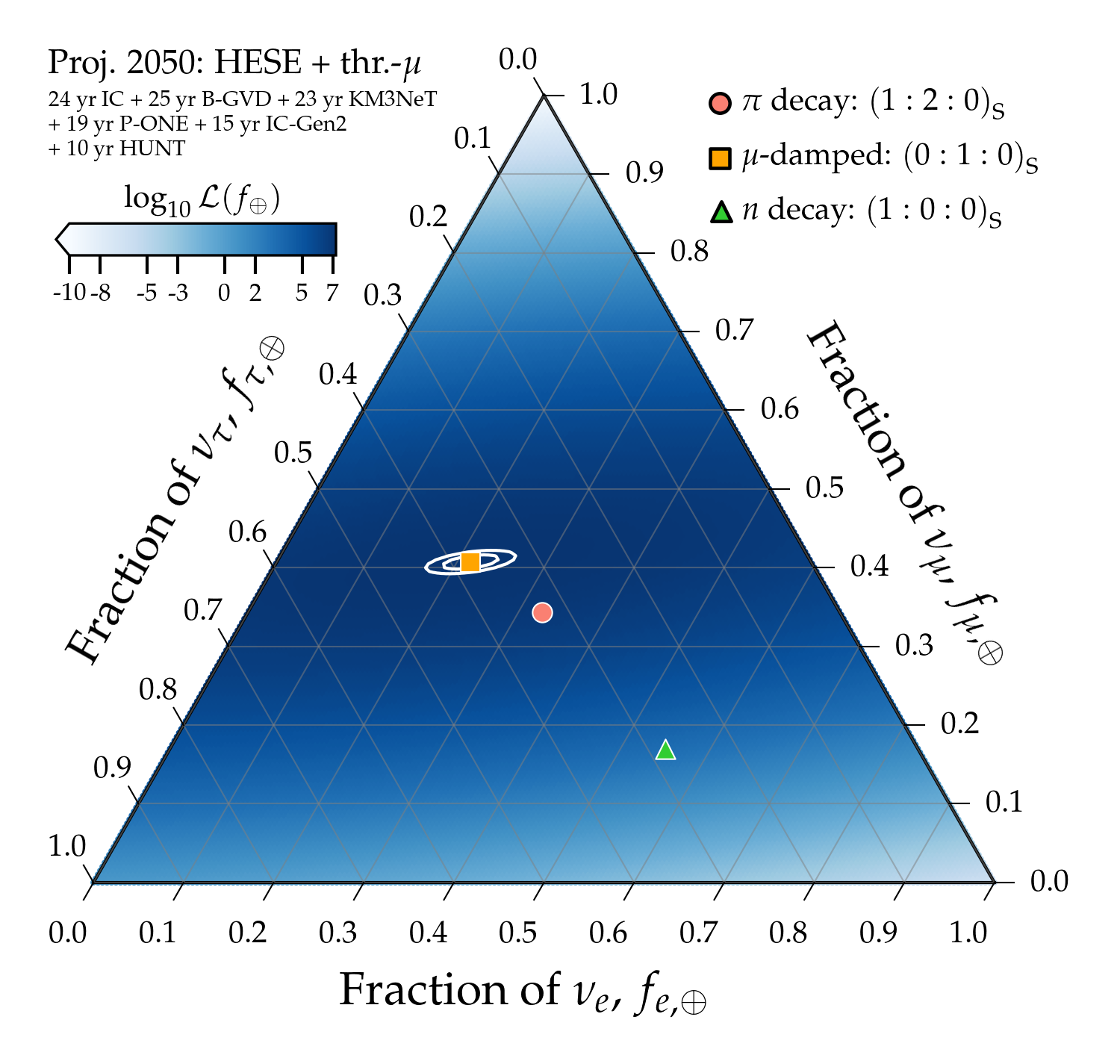}
    \end{minipage}
    \vspace*{-0.5cm}
    \caption{\textbf{\textit{Forecasts of experimental likelihood of flavor-composition measurements}}. Same as \figu{flavor_likelihood_mese}, but for forecasts using HESE plus through-going muons and assuming (\textit{left column}) neutrino production via pion decay, \ie, $(\frac{1}{3}, \frac{2}{3}, 0)_{\rm S}$, and (\textit{right column}) via muon-damped pion decay, \ie, $(0, 1, 0)_{\rm S}$  \textit{Top:} Using combined observations by multiple existing neutrino telescopes with km$^3$ size.  \textit{Center:} Combination of existing and future neutrino telescopes, including telescopes of 8~km$^3$.  \textit{Bottom:} Combination including one telescope with tens of km$^3$.}
    \label{fig:flavor_likelihood_future}
\end{figure*}

We consider four observation eras:
\begin{description}
 \item[Present]
  Figure~\ref{fig:flavor_likelihood_mese} shows the flavor-composition likelihood inferred by \Refe~\cite{Abbasi:2025fjc} from the 11.4-year IceCube MESE sample~\cite{IceCube:2025tgp}.  We use this likelihood to extract present-day measurements of the neutrino mixing parameters. (To derive our results, we use a close analytical approximation to the results reported by the IceCube Collaboration~\cite{Abbasi:2025fjc}.)
 \item[Projected, km$^3$-scale telescopes]
  Figure~\ref{fig:flavor_likelihood_future}, top row, shows the year-2040  likelihood obtained by combining observations made by existing neutrino telescopes IceCube (29 years of exposure), Baikal-GVD (15 yr), and KM3NeT (13 yr), each with a volume of roughly 1~km$^3$.  
 \item[Projected, multi-km$^3$ telescopes]
  Figure~\ref{fig:flavor_likelihood_future}, central row, shows the year-2050 likelihood obtained by combining observations made by existing and future detectors of km-scale size---IceCube (24 yr), Baikal-GVD (25 yr), KM3NeT (23 yr), P-ONE (19 yr)---and of about 8-km$^3$---IceCube-Gen2 (15 yr) and either NEON or TRIDENT (10 yr).
 \item[Projected, one tens-of-km$^3$ telescope]
  Figure~\ref{fig:flavor_likelihood_future}, bottom row, shows the year-2050 likelihood obtained by combining observations made by existing and future detectors of km-scale size---IceCube (24 yr), Baikal-GVD (25 yr), KM3NeT (23 yr), P-ONE (19 yr)---of about 8-km$^3$---IceCube-Gen2 (15 yr)---and of 30~km$^3$---HUNT (10 yr).
\end{description}

\section{Detailed results}
\label{app:detailed_results}

\renewcommand{\theequation}{B\arabic{equation}}
\renewcommand{\thefigure}{B\arabic{figure}}
\renewcommand{\thetable}{B\arabic{table}}
\setcounter{figure}{0} 
\setcounter{table}{0} 

\begin{table*}[t!]
 \begin{ruledtabular}
    \centering
    \caption{\label{tab:results_frequentist_full}\textbf{\textit{Frequentist sensitivity to neutrino mixing parameters from high-energy astrophysical neutrinos.}} Expanded version of Table~\ref{tab:results} in the main text. All results shown are obtained by constraining one parameter at a time while keeping the others controlled via pull terms informed by present (NuFIT 6.1~\cite{Esteban:2024eli}) and future~\cite{Song:2020nfh} sub-TeV global oscillation fits.  We assume no $\nu_\tau$ production (\ie, $f_{\tau, {\rm S}} = 0$). Figures~\ref{fig:2d_profiled_likelihood_pion_2040} and \ref{fig:2d_profiled_likelihood_pion_2050} show joint parameter distributions. The values are stacked: for each parameter, the top line shows the best fit $\pm$ $1\sigma$, and the bottom line (in brackets) shows the $3\sigma$ range, profiled over all the other parameters. See the main text and Appendix~\ref{app:detailed_results} for details.}
    \renewcommand{\arraystretch}{1.3}
    \setlength{\tabcolsep}{4pt} 
    \footnotesize
    \begin{tabular}{l c c c c c c c c}
        \multirow{3}{*}{Parameter} & \multirow{3}{*}{\makecell{Present \\ (IC MESE \\ 11.4 yr)}} & \multicolumn{6}{c}{Future (multi-detector projections using HESE plus through-going muons)} & \multirow{3}{*}{\makecell{Global \\ ($<$ TeV) \\ NuFIT 6.1}} \\
        \cline{3-8}
         & & \multicolumn{2}{c}{Only km$^3$-scale telescopes\footnotemark[1]} & \multicolumn{2}{c}{Plus multi-km$^3$ telescopes\footnotemark[2]} & \multicolumn{2}{c}{Plus tens-of-km$^3$ telescope\footnotemark[3]} & \\
        \cline{3-4} \cline{5-6} \cline{7-8}
         & & $\pi$ decay & $\mu$-damped & $\pi$ decay & $\mu$-damped & $\pi$ decay & $\mu$-damped & \\
        \hline
        \multirow{2}{*}{ $\sin^2 \theta_{12}$ } & $0.000_{-0.000}^{+1.000}$ & $0.291_{-0.291}^{+0.709}$ & $0.224_{-0.224}^{+0.128}$ & $0.279_{-0.279}^{+0.721}$ & $0.224_{-0.224}^{+0.104}$ & $0.290_{-0.290}^{+0.710}$ & $0.224_{-0.224}^{+0.102}$ & $0.303_{-0.012}^{+0.012}$ \\
         & $[0.000, 1.000]$ & $[0.000, 1.000]$ & $[0.000, 1.000]$ & $[0.000, 1.000]$ & $[0.000, 1.000]$ & $[0.000, 1.000]$ & $[0.000, 1.000]$ & $[0.270, 0.341]$ \\
        \cline{2-9}
        \multirow{2}{*}{ $\theta_{12}~[^\circ]$ } & $0.00_{-0.00}^{+90.00}$ & $32.66_{-32.66}^{+57.34}$ & $28.26_{-28.26}^{+8.13}$ & $31.86_{-31.86}^{+58.14}$ & $28.26_{-28.26}^{+6.68}$ & $32.59_{-32.59}^{+57.41}$ & $28.26_{-28.26}^{+6.57}$ & $33.40_{-0.75}^{+0.74}$ \\
         & $[0.00, 90.00]$ & $[0.00, 90.00]$ & $[0.00, 90.00]$ & $[0.00, 90.00]$ & $[0.00, 90.00]$ & $[0.00, 90.00]$ & $[0.00, 90.00]$ & $[31.31, 35.73]$ \\
        \hline
        \multirow{2}{*}{ $\sin^2 \theta_{23}$ } & $0.382_{-0.242}^{+0.618}$ & $0.470_{-0.153}^{+0.530}$ & $0.470_{-0.065}^{+0.293}$ & $0.471_{-0.088}^{+0.391}$ & $0.470_{-0.025}^{+0.144}$ & $0.471_{-0.076}^{+0.350}$ & $0.470_{-0.022}^{+0.127}$ & $0.451_{-0.016}^{+0.019}$ \\
         & $[0.000, 1.000]$ & $[0.166, 1.000]$ & $[0.139, 1.000]$ & $[0.295, 1.000]$ & $[0.354, 0.801]$ & $[0.318, 1.000]$ & $[0.374, 0.758]$ & $[0.408, 0.603]$ \\
        \cline{2-9}
        \multirow{2}{*}{ $\theta_{23}~[^\circ]$ } & $38.16_{-16.19}^{+51.84}$ & $43.26_{-8.99}^{+46.74}$ & $43.29_{-3.74}^{+17.60}$ & $43.32_{-5.13}^{+24.89}$ & $43.29_{-1.43}^{+8.29}$ & $43.32_{-4.38}^{+21.61}$ & $43.29_{-1.28}^{+7.33}$ & $42.19_{-0.92}^{+1.09}$ \\
         & $[0.00, 90.00]$ & $[24.06, 90.00]$ & $[21.91, 90.00]$ & $[32.89, 90.00]$ & $[36.54, 63.50]$ & $[34.31, 90.00]$ & $[37.70, 60.51]$ & $[39.70, 50.94]$ \\
        \hline
        \multirow{2}{*}{ $\sin^2 \theta_{13}$ } & $0.000_{-0.000}^{+0.714}$ & $0.013_{-0.013}^{+0.602}$ & $0.023_{-0.019}^{+0.134}$ & $0.014_{-0.014}^{+0.499}$ & $0.023_{-0.011}^{+0.037}$ & $0.014_{-0.014}^{+0.475}$ & $0.023_{-0.009}^{+0.030}$ & $0.02248_{-0.00059}^{+0.00055}$ \\
         & $[0.000, 1.000]$ & $[0.000, 0.899]$ & $[0.000, 0.866]$ & $[0.000, 0.664]$ & $[0.001, 0.203]$ & $[0.000, 0.616]$ & $[0.002, 0.156]$ & $[0.02064, 0.02418]$ \\
        \cline{2-9}
        \multirow{2}{*}{ $\theta_{13}~[^\circ]$ } & $0.00_{-0.00}^{+57.66}$ & $6.55_{-6.55}^{+45.08}$ & $8.63_{-5.19}^{+14.71}$ & $6.67_{-6.67}^{+39.05}$ & $8.63_{-2.37}^{+5.45}$ & $6.67_{-6.67}^{+37.64}$ & $8.63_{-2.06}^{+4.57}$ & $8.54_{-0.12}^{+0.11}$ \\
         & $[0.00, 90.00]$ & $[0.00, 71.52]$ & $[0.00, 68.53]$ & $[0.00, 54.58]$ & $[1.81, 26.74]$ & $[0.00, 51.69]$ & $[2.69, 23.30]$ & $[8.19, 8.90]$ \\
        \hline
        \multirow{2}{*}{ $\delta_{\rm CP}~[^\circ]$ } & $164_{-164}^{+196}$ & $132_{-132}^{+228}$ & $148_{-28}^{+93}$ & $147_{-147}^{+213}$ & $148_{-14}^{+77}$ & $147_{-147}^{+213}$ & $148_{-12}^{+76}$ & $232_{-26}^{+36}$ \\
         & $[0, 360]$ & $[0, 360]$ & $[57, 303]$ & $[0, 360]$ & $[111, 249]$ & $[0, 360]$ & $[116, 244]$ & $[133, 368]$ \\
        \hline
        \multirow{2}{*}{ $f_{e, {\rm S}}$ } & $0.043_{-0.043}^{+0.492}$ & $0.334_{-0.053}^{+0.053}$ & $0.000_{-0.000}^{+0.061}$ & $0.334_{-0.026}^{+0.026}$ & $0.000_{-0.000}^{+0.028}$ & $0.334_{-0.020}^{+0.019}$ & $0.000_{-0.000}^{+0.026}$ & \makecell{$\cdots$} \\
         & $[0.000, 1.000]$ & $[0.174, 0.491]$ & $[0.000, 0.185]$ & $[0.256, 0.409]$ & $[0.000, 0.084]$ & $[0.273, 0.389]$ & $[0.000, 0.076]$ & \makecell{$\cdots$} \\
    \end{tabular}
    \footnotetext[1]{Projections for 2040 using 29~yr of IceCube + 15~yr of Baikal-GVD + 13~yr of KM3NeT.}
    \footnotetext[2]{Projections for 2050 using 24~yr of IceCube + 25~yr of Baikal-GVD + 23~yr of KM3NeT + 19~yr of P-ONE + 15~yr of IceCube-Gen2 + 10~yr of NEON or TRIDENT.}
    \footnotetext[3]{Projections for 2050 using 24~yr of IceCube + 25~yr of Baikal-GVD + 23~yr of KM3NeT + 19~yr of P-ONE + 15~yr of IceCube-Gen2 + 10~yr of HUNT.}
 \end{ruledtabular}
\end{table*}

This appendix presents comprehensive numerical results and joint parameter distributions for our frequentist analysis, expanding on the condensed results shown in the main text.

Table~\ref{tab:results_frequentist_full} shows detailed numerical results for the constraints on the mixing parameters, expanding on  Table~\ref{tab:results} in the main text.  Table~\ref{tab:results_frequentist_full} reveals an important prospect: achieving a measurement error of 10\% or better will require combining measurements by existing km$^3$-scale neutrino telescopes and upcoming multi-km$^3$ ones.

\begin{figure*}[t!]
 \centering
 \includegraphics[width=\textwidth]{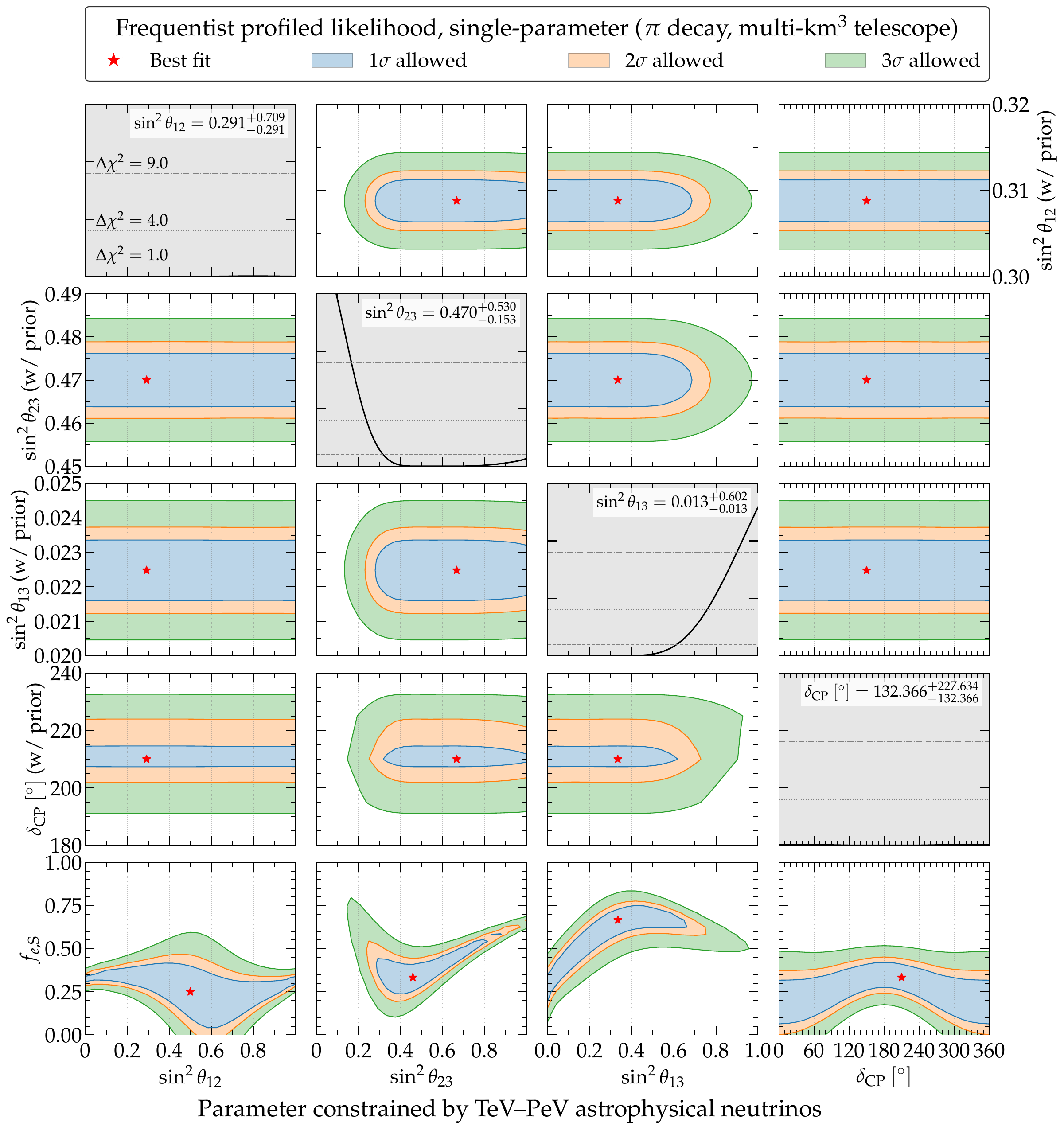}
 \caption{\textbf{\textit{Future pairwise frequentist profiled likelihood using multi-km$^3$ detectors.}} The flavor-measurement likelihood uses the combined exposure of IceCube, Baikal-GVD, and KM3NeT to the year 2040, assuming neutrino production via full pion decay and $f_{\tau, {\rm S}} = 0$. See Appendix~\ref{app:flavor_likelihood} for detector details.}
 \label{fig:2d_profiled_likelihood_pion_2040}
\end{figure*}

\begin{figure*}[t!]
 \centering
 \includegraphics[width=\textwidth]{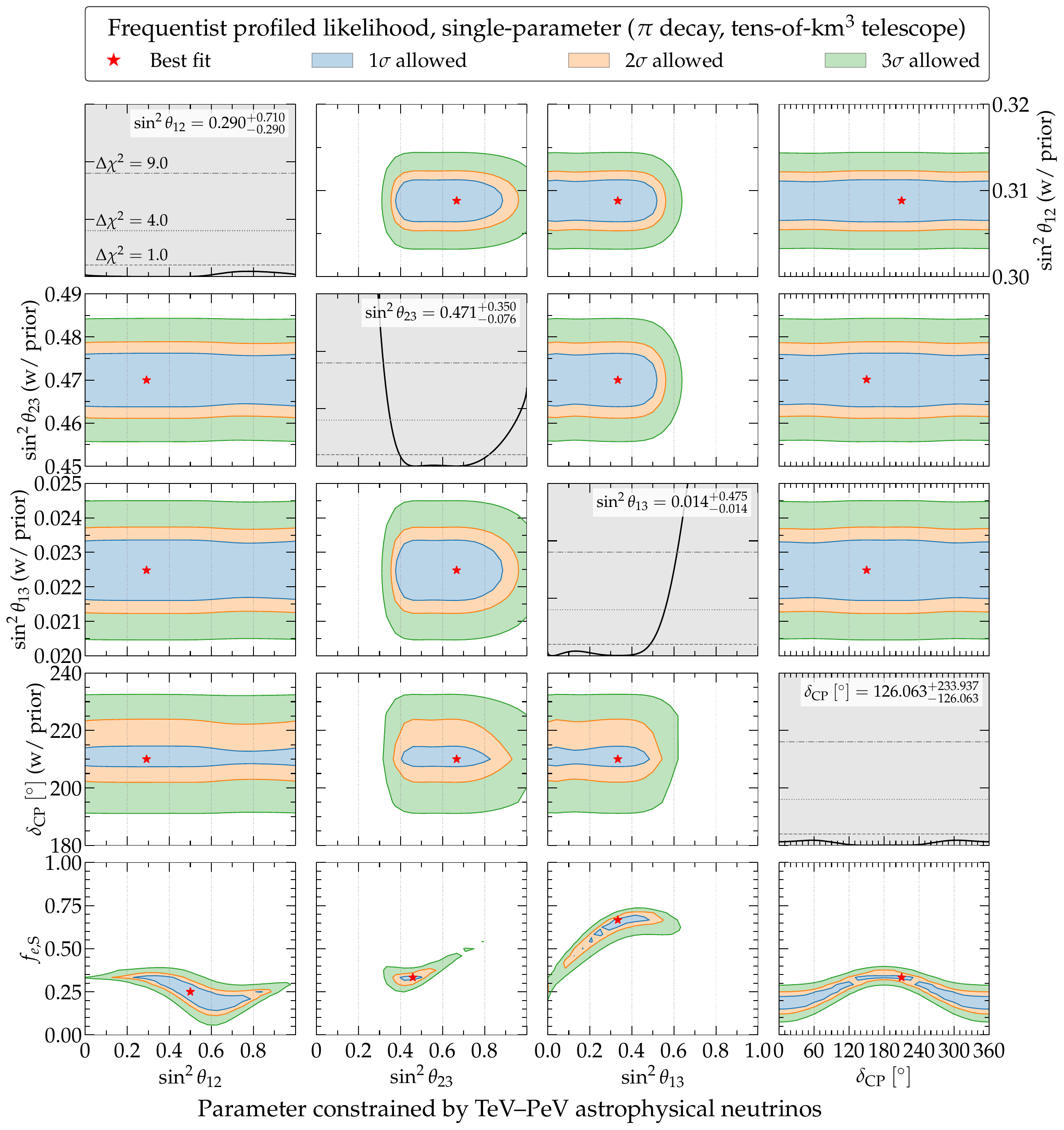}
 \caption{\textbf{\textit{Future pairwise frequentist profiled likelihood using multi-km$^3$ detectors.}} Same as \figu{2d_profiled_likelihood_pion_2040}, but the flavor-measurement likelihood uses the combined exposure of IceCube, Baikal-GVD, KM3NeT, P-ONE, IceCube-Gen2, and HUNT to the year 2050, assuming neutrino production via full pion decay  and $f_{\tau, {\rm S}} = 0$.  See Appendix~\ref{app:flavor_likelihood} for detector details.}
 \label{fig:2d_profiled_likelihood_pion_2050}
\end{figure*}

Figures~\ref{fig:2d_profiled_likelihood_pion_2040} and \ref{fig:2d_profiled_likelihood_pion_2050} show projected pairwise profiled likelihood functions, under our single-parameter approach while keeping the others constrained with sub-TeV pull terms, using multi-km$^3$ neutrino telescopes and adding a tens-of-km$^3$ telescope, respectively.  The pairwise profiled likelihood is a straightforward generalization of the one-dimensional profiled likelihood defined in the main text, for two degrees of freedom instead of one.

\begin{figure}[t!]
 \centering
 \includegraphics[width=\columnwidth]{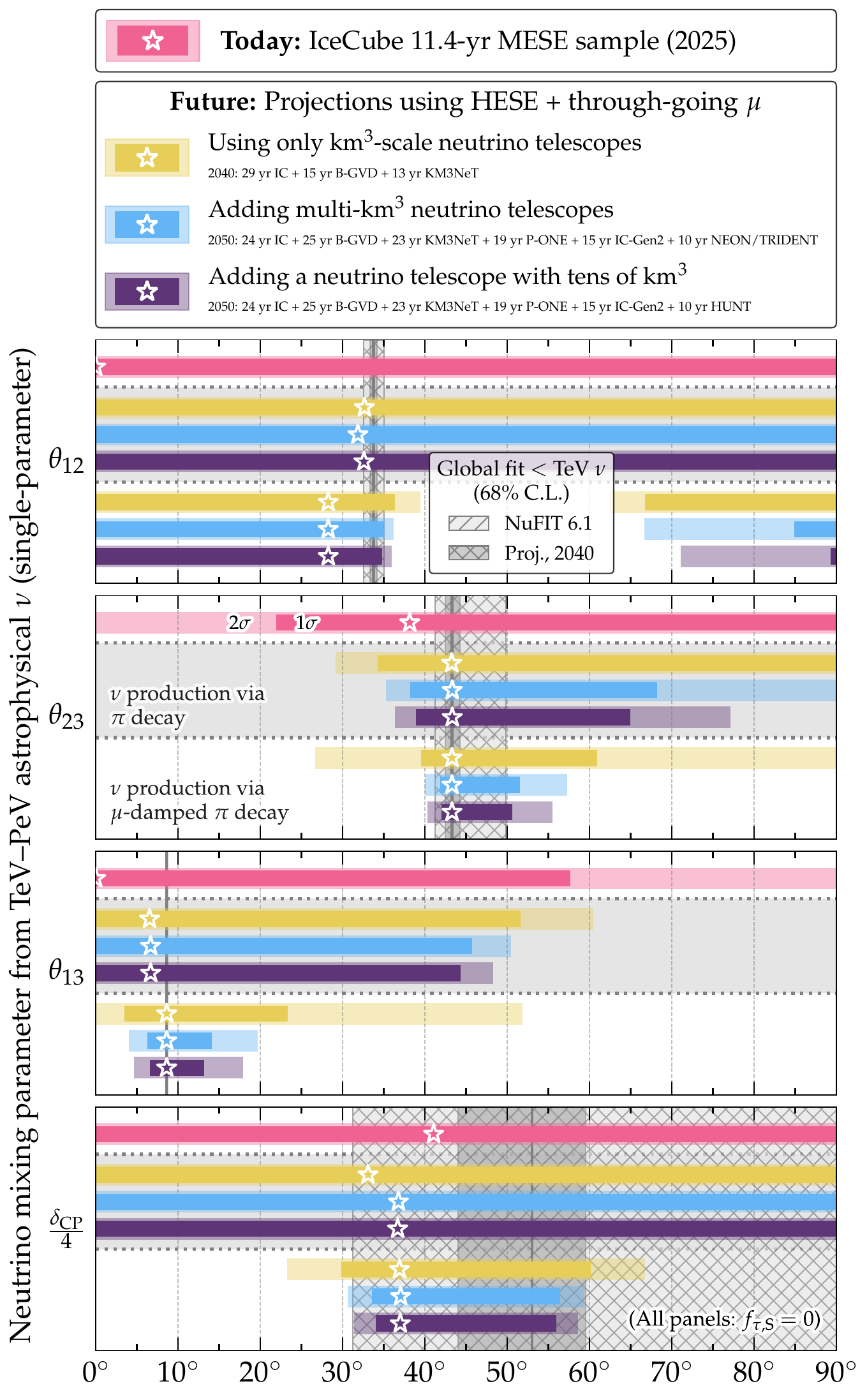}
 \vspace*{-0.5cm}
 \caption{\textbf{\textit{Prospects for constraining neutrino mixing parameters with high-energy astrophysical neutrinos.}} Same as \figu{results_main} in the main text, but allowing for $\nu_\tau$ production at the astrophysical sources, \ie, allowing $f_{\tau, {\rm S}} $ to vary between 0 and 1.  The results do not change significantly compared to our main results obtained with $f_{\tau, {\rm S}} = 0$.}
 \vspace*{-0.5cm}
 \label{fig:results_frequentist_summary_ftS_free}
\end{figure}

Figure~\ref{fig:results_frequentist_summary_ftS_free} shows that the parameter constraints obtained allowing for $\nu_\tau$ production at the sources are not significantly weaker than our main ones, obtained assuming $f_{\tau, {\rm S}} = 0$.  This is to be expected, since the allowed region of flavor composition at Earth grows only slightly when allowing for $f_{\tau, {\rm S}} \neq 0$~\cite{Bustamante:2015waa}.


\section{Bayesian results}
\label{app:bayesian_results}

\renewcommand{\theequation}{C\arabic{equation}}
\renewcommand{\thefigure}{C\arabic{figure}}
\renewcommand{\thetable}{C\arabic{table}}
\setcounter{figure}{0} 
\setcounter{table}{0} 

\begin{table*}[t!]
 \begin{ruledtabular}
  \centering
  \caption{\label{tab:results_bayesian}\textbf{\textit{Bayesian sensitivity to neutrino mixing parameters from high-energy astrophysical neutrinos.}} Similar to Table~\ref{tab:results_frequentist_full}, but obtained using the Bayesian prescription and priors described in Appendix~\ref{app:bayesian_results}. We assume no $\nu_\tau$ production (\ie, $f_{\tau, {\rm S}} = 0$). The values are stacked: for each parameter, the top line shows the best fit (``B.f.'')  $\pm$ 68\% C.L., and the bottom line (in brackets) shows the 95\% C.L. range. Best-fit values are the medians of the one-dimensional parameter posteriors.  The Bayes factor, $\mathcal{B}$, represents the strength of the evidence in favor of three-flavor mixing.} 
  \renewcommand{\arraystretch}{1.3}
  \setlength{\tabcolsep}{5pt} 
  \footnotesize
  \begin{tabular}{cccccccc}
   \multirow{4}{*}{Parameter} & 
   \multicolumn{3}{c}{Conservative: All mixing parameters free} & 
   \multicolumn{3}{c}{Aggressive: Single-parameter constraints} &
   \multirow{4}{*}{\shortstack{Global ($<$ TeV) \\ NuFIT 6.1 \\ \scriptsize (B.f. $\pm$ 68\% C.L. \\ \scriptsize / [$3\sigma$])}} \\
   & \multicolumn{3}{c}{\small (B.f. $\pm$ 68\% C.L. / [95\% C.L.])} & \multicolumn{3}{c}{\small (B.f. $\pm$ 68\% C.L. / [$95\%$~C.L.])} & 
    \\
   \cline{2-4} \cline{5-7} 
   & \multirow{2}{*}{\shortstack{Present \\ ($\mathcal{B} \approx 16$)}} & \multicolumn{2}{c}{Future\footnotemark[1] ($\mathcal{B} \sim 10^{56}, 10^{61}$)} & \multirow{2}{*}{Present} & \multicolumn{2}{c}{Future\footnotemark[1]} & \\
   \cline{3-4} \cline{6-7}
   & & $\pi$ decay & $\mu$-damped & & $\pi$ decay & $\mu$-damped & \\
   \hline
   \multirow{2}{*}{$\sin^2 \theta_{12}$}  
     & $0.503_{-0.312}^{+0.304}$ & $0.498_{-0.283}^{+0.284}$ & $0.506_{-0.405}^{+0.397}$
     & $0.507_{-0.278}^{+0.279}$ & $0.531_{-0.243}^{+0.210}$ & $0.209_{-0.123}^{+0.088}$
     & $0.303_{-0.012}^{+0.012}$ \\
     & [0.033, 0.970] & [0.037, 0.962] & [0.013, 0.986]
     & [0.043, 0.957] & [0.071, 0.935] & [0.018, 0.976]
     & [0.270, 0.341] \\ \cline{2-8}
     
   \multirow{2}{*}{$\theta_{12}~[^\circ]$} 
     & $45.18_{-19.26}^{+18.77}$ & $44.89_{-17.26}^{+17.28}$ & $45.34_{-26.81}^{+26.51}$
     & $45.40_{-16.81}^{+17.04}$ & $46.78_{-14.32}^{+12.63}$ & $27.20_{-10.15}^{+5.82}$
     & $33.40_{-0.75}^{+0.74}$ \\
     & [10.47, 80.03] & [11.09, 78.76] & [6.55, 83.20]
     & [11.97, 78.03] & [15.45, 75.23] & [7.71, 81.09]
     & [31.63, 35.95] \\ \cline{2-8}
     
   \multirow{2}{*}{$\sin^2 \theta_{23}$}  
     & $0.509_{-0.326}^{+0.313}$ & $0.497_{-0.224}^{+0.220}$ & $0.452_{-0.230}^{+0.203}$
     & $0.454_{-0.255}^{+0.304}$ & $0.541_{-0.113}^{+0.172}$ & $0.535_{-0.049}^{+0.067}$
     & $0.451_{-0.016}^{+0.019}$ \\
     & [0.031, 0.971] & [0.049, 0.923] & [0.044, 0.920]
     & [0.053, 0.954] & [0.370, 0.863] & [0.451, 0.670]
     & [0.435, 0.585] \\ \cline{2-8}
     
   \multirow{2}{*}{$\theta_{23}~[^\circ]$} 
     & $45.52_{-20.19}^{+19.53}$ & $44.83_{-13.33}^{+13.03}$ & $42.25_{-14.14}^{+11.78}$
     & $42.36_{-15.87}^{+18.17}$ & $47.35_{-6.49}^{+10.26}$ & $47.01_{-2.81}^{+3.88}$
     & $42.19_{-0.92}^{+1.09}$ \\
     & [10.14, 80.20] & [12.79, 73.89] & [12.11, 73.57]
     & [13.31, 77.62] & [37.47, 68.28] & [42.19, 54.94]
     & [39.70, 50.94] \\ \cline{2-8}
     
   \multirow{2}{*}{$\sin^2 \theta_{13}$}  
     & $0.427_{-0.256}^{+0.296}$ & $0.390_{-0.193}^{+0.253}$ & $0.761_{-0.638}^{+0.151}$
     & $0.389_{-0.250}^{+0.293}$ & $0.247_{-0.172}^{+0.167}$ & $0.047_{-0.020}^{+0.033}$
     & $0.02248_{-0.00059}^{+0.00055}$ \\
     & [0.021, 0.932] & [0.037, 0.888] & [0.015, 0.983]
     & [0.025, 0.915] & [0.011, 0.513] & [0.015, 0.128]
     & [0.02064, 0.02418] \\ \cline{2-8}
     
   \multirow{2}{*}{$\theta_{13}~[^\circ]$} 
     & $40.80_{-16.38}^{+17.44}$ & $38.65_{-12.30}^{+14.66}$ & $60.73_{-40.20}^{+12.01}$
     & $38.59_{-16.70}^{+17.09}$ & $29.80_{-13.91}^{+10.25}$ & $12.52_{-3.06}^{+3.91}$
     & $8.62_{-0.11}^{+0.11}$ \\
     & [8.33, 74.88] & [11.09, 70.45] & [7.03, 82.51]
     & [9.10, 73.05] & [6.02, 45.74] & [7.03, 20.96]
     & [8.26, 8.95] \\ \cline{2-8}
     
   \multirow{2}{*}{$\delta_{\rm CP}~[^\circ]$} 
     & $185.34_{-124.13}^{+114.63}$ & $182.66_{-110.08}^{+104.12}$ & $178.01_{-106.16}^{+111.13}$
     & $179.15_{-120.30}^{+122.81}$ & $179.15_{-131.21}^{+131.38}$ & $180.16_{-24.93}^{+25.27}$
     & $212_{-36}^{+26}$ \\
     & [9.79, 350.07] & [18.79, 343.44] & [15.39, 344.01]
     & [10.22, 350.84] & [6.98, 352.43] & [138.05, 222.47]
     & [125, 365] \\ 
     
   \hline 
   \multirow{2}{*}{$f_{e, {\rm S}}$} 
     & $0.446_{-0.265}^{+0.304}$ & $0.382_{-0.215}^{+0.298}$ & $0.143_{-0.091}^{+0.063}$
     & $0.339_{-0.214}^{+0.293}$ & $0.332_{-0.020}^{+0.019}$ & $0.017_{-0.012}^{+0.019}$
     & $\cdots$ \\
     & [0.039, 0.951] & [0.028, 0.920] & [0.010, 0.250]
     & [0.024, 0.886] & [0.293, 0.368] & [0.001, 0.056]
     & $\cdots$ \\
  \end{tabular}
  \footnotetext[1]{Multi-detector projections of HESE plus through-going muons using 24~yr of IceCube + 25~yr of Baikal-GVD + 23~yr of KM3NeT + 19~yr of P-ONE + 15~yr of IceCube-Gen2 + 10~yr of HUNT.}
 \end{ruledtabular}
\end{table*}

This appendix presents complementary Bayesian  constraints on the neutrino mixing parameters, providing an alternative statistical perspective to the frequentist analysis shown in the main text and Appendix~\ref{app:detailed_results}. 

We compute Bayesian posterior probability densities to assess the sensitivity to mixing parameters, $\boldsymbol{\theta} = (\theta_{12}, \theta_{23}, \theta_{13}, \delta_{\rm CP})$, \ie,
\begin{equation*}
 \label{equ:posterior}
 \mathcal{P}(\boldsymbol{\theta}, f_{e, {\rm S}})
 =
 \pi(\boldsymbol{\theta}) 
 \pi(f_{e, {\rm S}})
 \mathcal{L}
 [
 f_{e, \oplus}(f_{e, {\rm S}}, \boldsymbol{\theta}), 
 f_{\mu, \oplus}(f_{e, {\rm S}}, \boldsymbol{\theta})
 ] \;,
\end{equation*}
where $\pi(\boldsymbol{\theta})$ is the prior probability distribution on the mixing parameters, $\pi(f_{e, {\rm S}})$ is uniform in $[0, 1]$, and $\mathcal{L}$ is the likelihood of flavor-composition measurements, present or future, as in the frequentist approach.  We assume no $\nu_\tau$ production (\ie, $f_{\tau, {\rm S}} = 0$). 

We produce Bayesian results under two approaches. First, we follow the same approach as in the frequentist analysis and measure a single parameter at a time while marginalizing over all the other parameters, using for them tight sub-TeV NuFIT-6.1 or projected priors. Second, separately, we measure all of the mixing parameters and $f_{e, {\rm S}}$ simultaneously, using uninformed priors for all of them.  When measurimg  $\sin^2(\theta_{ij})$, we use for it a uniform prior in $[0, 1]$; when measuring $\delta_{\rm CP}$, we use for it a uniform prior in $[0, 2\pi]$.  

\begin{figure*}[t]
 \centering
 \begin{minipage}{\columnwidth}
  \centering
  \includegraphics[width=\columnwidth]{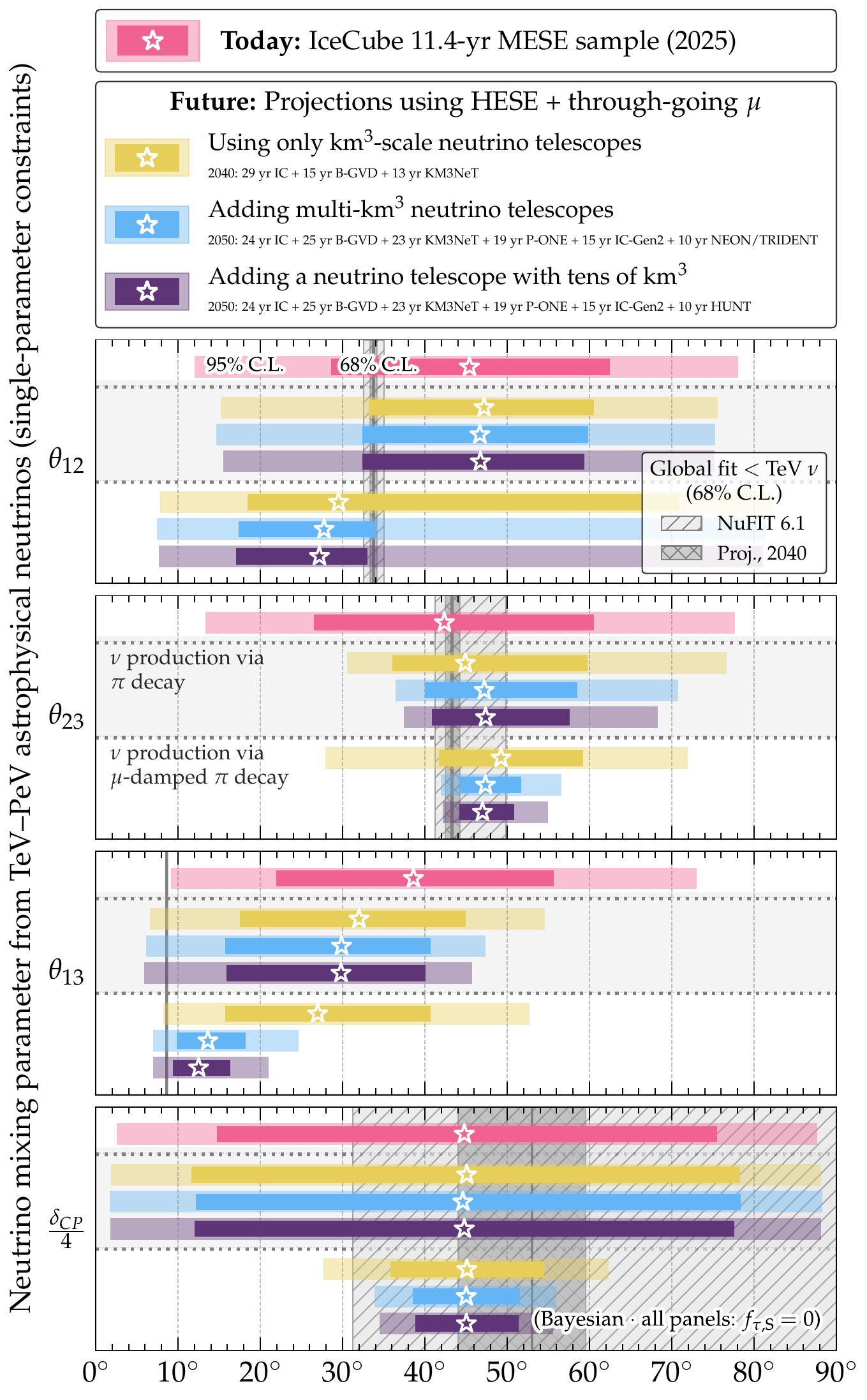}
  \label{fig:results_bayesian_aggressive}
 \end{minipage}%
 \hfill 
 \begin{minipage}{\columnwidth}
  \centering
  \includegraphics[width=\columnwidth]{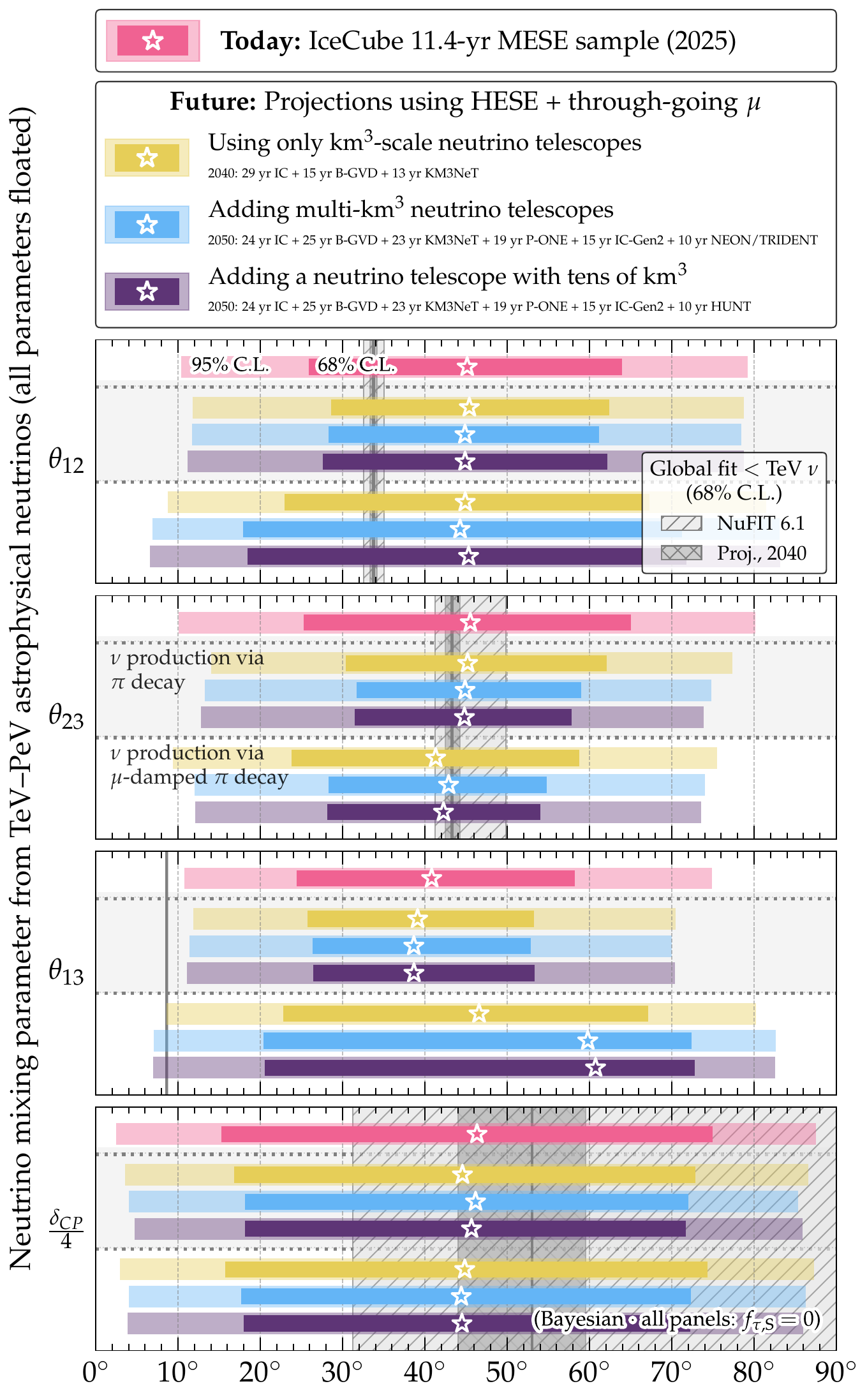}
  \label{fig:results_bayesian_conservative}
 \end{minipage}
 \vspace*{-0.5cm}
 \caption{\textbf{\textit{Bayesian Prospects for constraining neutrino mixing parameters with high-energy astrophysical neutrinos.}} Similar to \figu{results_main} in the main text, but with intervals obtained via a Bayesian analysis instead of the frequentist analysis of our main results.  Results are obtained by constraining a single parameter at a time (\textit{left panel}), as in the main text, and all of them simultaneously (\textit{right panel}).
 \vspace*{-0.5cm}}
 \label{fig:results_bayesian_1d}
\end{figure*}

Table~\ref{tab:results_bayesian} and \figu{results_bayesian_1d} show our Bayesian results. Present constraints from the 11.4-year IceCube MESE sample~\cite{Abbasi:2025fjc} show wide posteriors spanning most of the allowed range, indicating insufficient constraining power.  Future multi-telescope projections show improved sensitivity, particularly for $\theta_{23}$, as in the frequentist case.  Quantitatively, the Bayes factor $\mathcal{B}$ comparing the statistical evidence (\ie, the fully integrated posterior) with flavor mixing vs.~without it is about 16 today, lending strong---but not decisive~\cite{Jeffreys:1939xee}---evidence for three-flavor mixing.  In our projections, $\mathcal{B}$ grows to $10^{56}$--$10^{61}$, depending on the neutrino production assumed, lending incontrovertible evidence for three-flavor mixing.

Broadly stated, the frequentist and Bayesian results agree.  Yet, there are important differences that make us select the former as our main results, as we explain next.

The frequentist constraints are more conservative than the baseline Bayesian ones.  This is clearest when comparing present-day results between Tables~\ref{tab:results_frequentist_full} and \ref{tab:results_bayesian}. While $1\sigma$ frequentist constraints span the entire allowed physical range of most parameters, representing complete lack of sensitivity, the $68\%$~C.L.~Bayesian constraints are narrower, representing nascent sensitivity.    

However, this is deceptive, and merely an artifact of our choice of priors.  This is conveyed by the fact that in Table~\ref{tab:results_bayesian} the best-fit values (\ie, the median) for $\sin^2 \theta_{12}$, $\sin^2 \theta_{23}$, and $\delta_{\rm CP}$ are merely the mid-points of their uniform priors, \ie, 0.5 and $180^\circ$, respectively.  (There is marginally more sensitivity to $\sin^2 \theta_{13}$.)  In turn, this is a reflection of the large errors in the present-day flavor-composition measurements from the 11.4-year IceCube MESE sample.  Using different priors on the  parameters, \eg, log-uniform ones that can more efficiently sample their lowest allowed values, widens their posteriors and washes away any present Bayesian sensitivity. 

This contrast between the frequentist and Bayesian approaches is mitigated in our forecasts. In them, their constraints become more alike as a result of the improved flavor-composition measurements, reducing the impact of our choice of priors on the Bayesian posteriors.  This justifies why, to be conservative in our claims and consistent between our present and projected results, we choose to show frequentist constraints rather than Bayesian ones as our main results.  In the future, once flavor-composition measurements improve, the choice between these approaches will become less impactful.

\end{document}